\documentclass[final,3p,times]{elsarticle}

\usepackage{amsfonts}
\usepackage{amssymb}
\usepackage{amsmath}
\usepackage{graphics}
\usepackage{epsfig}
\usepackage{amsthm}
\usepackage{textcomp}
\usepackage{color}
\usepackage{float}
\usepackage{multirow}
\usepackage{graphicx}
\usepackage{subfigure}
\usepackage{longtable}
\usepackage{rotating}
\usepackage{mathrsfs}

\newtheorem{expe}{Experiment}

\journal{Physica A}
\begin{document}
\begin{frontmatter}
		
		
        \cortext[cor1]{Corresponding author}
		\title{An effective rumor-containing strategy}
		
		
		\author[label1]{Cheng Pan}
		\ead{mrpanc520@gmail.com}
		
		\author[label2]{Lu-Xing Yang}
		\ead{ylx910920@gmail.com}
		
		\author[label1]{Xiaofan Yang\corref{cor1}}
		\ead{xfyang1964@gmail.com}
		
		\author[label1]{Yingbo Wu}
		\ead{wyb@cqu.edu.cn}
		
		\author[label3]{Yuan Yan Tang}
		\ead{yytang@umac.mo}
		
		\address[label1]{School of Software Engineering, Chongqing University, Chongqing, 400044, China}
		
		\address[label2]{School of Information Technology, Deakin University, Melbourne, 3125, Australia}
		
		\address[label3]{Department of Computer and Information Science, The University of Macau, Macau}

\begin{abstract}

False rumors can lead to huge economic losses or/and social instability. Hence, mitigating the impact of bogus rumors is of primary importance. This paper focuses on the problem of how to suppress a false rumor by use of the truth. Based on a set of rational hypotheses and a novel rumor-truth mixed spreading model, the effectiveness and cost of a rumor-containing strategy are quantified, respectively. On this basis, the original problem is modeled as a constrained optimization problem (the RC model), in which the independent variable and the objective function represent a rumor-containing strategy and the effectiveness of a rumor-containing strategy, respectively. The goal of the optimization problem is to find the most effective rumor-containing strategy subject to a limited rumor-containing budget. Some optimal rumor-containing strategies are given by solving their respective RC models. The influence of different factors on the highest cost effectiveness of a RC model is illuminated through computer experiments. The results obtained are instructive to develop effective rumor-containing strategies. 

\end{abstract}

\end{frontmatter}
\section{Introduction}
	
Rumors are a common form of social interactions. A rumor is dispersed for achieving specific purpose, especially when a major public event occur and people do not have exact information and knowledge about the event \cite{Sunstein2009, Dubois2011, Serrano2015}. The rapidly popularized online social networks (OSNs) greatly enhance the speed and enlarge the extent of rumor spreading \cite{LiuDC2011, Doerr2012}. However, most false rumors can inflict economic losses or disrupt social order \cite{Thomas2007}. To exemplify, Syrian hackers once broke into the twitter account of Associated Press (AP) and dispersed the rumor that explosions at White House had injured Obama, leading to a loss of 10 billion US dollars before the rumor was clarified \cite{Peter2013}. Therefore, mitigating the impact of bogus rumors is of primary importance. \cite{Budak2011}. 

To effectively inhibit rumors, we have to gain insight into the laws governing rumor spreading. For this purpose, we need to establish and study appropriate rumor spreading models. The seminal work by Daley and Kendal \cite{Daley1965} introduced the first rumor spreading model, in which a homogeneously mixed population is divided into three groups: ignorants who are unaware of the rumor, spreaders who are aware of the rumor and spread it, and stiflers who are aware of the rumor but do not spread it. See Ref. \cite{Daley2001} for a popular introduction of the model. Since then, a multitude of rumor spreading models based on homogeneous networks have been proposed \cite{ZhaoLJ2011a, ZhaoLJ2012a, HuangWT2011, HuoLA2012a, ZhaoLJ2015, Afassinou2014, HuoLA2016, HuoLA2015, XuJP2016}. 

Empirical studies that started about two decades ago fully indicate that most realistic OSNs are heterogeneous rather than homogeneous \cite{Albert2002, Ebel2002}. In the past decade, therefore, much efforts were focused on rumor spreading models based on complex networks \cite{Nekovee2007, ZhouJ2007, Roshani2012, Singh2012, Naimi2013, ZhaoLJ2013a, ZhaoLJ2013b, WangYQ2013, ZanYL2014, HeZB2015, XiaLL2015, QiuXY2016, HeZB2017, LiuQM2017, LiD2017, Oliveros2017}. However, it is uncertain whether these models accurately characterize actual rumor spreading processes, because the models are derived through a series of approximations and do not perfectly accommodate the spreading network \cite{Satorras2001, Satorras2002, Satorras2015}. 

In 2009, Van Mieghem et al. \cite{Mieghem2009} proposed an elegant individual-level susceptible-infected-susceptible (SIS) epidemic model, in which the time evolution of the state of each individual is characterized by a separate differential equation. Due to the exact modeling and the perfect accommodation of the network topology, this model characterizes the SIS epidemics more accurately as compared with all previous SIS models. Moreover, this model is analytically tractable, resulting in profound conclusions. In recent years, this individual-based modeling approach has been widely applied to areas such as epidemic spreading \cite{Mieghem2011, Sahneh2012, Sahneh2013}, malware propagation \cite{XuSH2012a, XuSH2012b, XuSH2014, YangLX2015, YangLX2017a, YangLX2017b, YangLX2018a, YangLX2018b}, cyber security \cite{XuSH2015, ZhengR2015, YangLX2017c, ZhengR2018, YangXF2018}, and message transmission \cite{YangLX2017d}. To our knowledge, to date the rumor-containing problem has not yet been studied under individual-level rumor spreading models.

Clarifying rumors by releasing truths is a common way to inhibit rumors \cite{WenS2014, WenS2015}. In this context, every individual may choose to believe the rumor or believe the truth or be uncertain. This paper focuses on the problem of how to inhibit a false rumor by use of the truth. Based on a novel individual-level rumor-truth mixed spreading model, the effectiveness and cost of a rumor-containing strategy are quantified. Thereby, the original problem is modeled as a constrained optimization problem (the \emph{rumor-containing (RC) model}), in which the independent variable and the objective function represent a rumor-containing strategy and the effectiveness of a rumor-containing strategy, respectively. The goal of the optimization problem is to find the most effective rumor-containing strategy subject to a limited rumor-containing budget. Some optimal rumor-containing strategies are given by solving their respective RC models. The influence of different factors on the highest cost effectiveness of a RC model is uncovered through computer simulations. These results are condusive to the design of practical rumor-containing strategies. To our knowledge, this is the first time the rumor-containing problem is treated in this way.

The subsequent materials of this work are organized in this fashion: Section 2 establishes the rumor-truth mixed spreading model; Sections 3 and 4 formulate and study the RC model, respectively; Section 5 examines the influence of different factors on the highest cost effectiveness; this work is closed by Section 6.

\section{The modeling of the rumor-truth mixed spreading process}
	
This paper focuses on the following problem:

\emph{The rumor-containing (RC) problem:} Find an effective strategy of containing a false rumor, provided the truth is released.

The key to solving the problem is to properly quantify the effectiveness of a rumor-containing strategy. For this purpose, we have to model and study the rumor-truth mixed spreading process. This section is dedicated to establishing the model. 

\subsection{The state of a population}
	
Consider a closed population of $N$ individuals labeled $1, \cdots, N$. Let $V = \{1, \cdots, N\}$. Suppose in the time horizon $[0, T]$ a false rumor is dispersed in the population through a rumor-spreading network $G_R = (V, E_R)$, where each node stands for an individual, and $(i,j) \in E_R$ if and only if person $i$ can deliver the rumor to person $j$. Let $\mathbf{A} =\left[a_{ij}\right]_{N \times N}$ denote the adjacency matrix of $G_R$, i.e., $a_{ij} = 1$ or 0 according as $(i, j) \in E_R$ or not. Suppose in the time horizon $[0, T]$ the truth against the rumor is circulated in the population through a truth-spreading network $G_T = (V, E_T)$, where $(i,j) \in E_T$ if and only if person $i$ can deliver the truth to person $j$. Let $\mathbf{B} =\left[b_{ij}\right]_{N \times N}$ denote the adjacency matrix of $G_T$. 
	
At any time in the time horizon $[0, T]$, every individual in the population is in one of three possible states: \emph{rumor-believing}, \emph{truth-believing}, and \emph{uncertain}. Let $X_i(t) = 0$, 1, and 2 denote the event that person $i$ is uncertain, rumor-believing, and truth-believing at time $t$, respectively. Then, the vector
\begin{equation}
  \mathbf{X}(t) = \left[X_1(t), X_2(t), \cdots, X_N(t)\right].
\end{equation}
represents the \emph{state} of the population at time $t$.

Let $U_i(t)$, $R_i(t)$, and $T_i(t)$ denote the probability of the event that person $i$ is uncertain, rumor-believing, and truth-believing at time $t$, respectively, i.e., 
\begin{equation}
  U_i(t) = \Pr\left\{X_i(t) = 0\right\}, \quad R_i(t) = \Pr\left\{X_i(t) = 1\right\}, \quad T_i(t) = \Pr\left\{X_i(t) = 2\right\}.
\end{equation}
As $U_i(t)+R_i(t)+T_i(t)\equiv1$, the vector
\begin{equation}
    \mathbf {E}(t) = \left[R_1(t),\cdots,R_N(t), T_1(t),\cdots, T_N(t)\right]^T
\end{equation}
represents the \emph{expected state} of the population at time $t$. Let 
\begin{equation}
  \mathbf{R}(t) = \left[R_1(t), \cdots, R_N(t)\right], \quad \mathbf{T}(t) = \left[T_1(t), \cdots, T_N(t)\right].
\end{equation}
Then, $\mathbf {E}(t) =\left[\mathbf{R}(t), \mathbf{T}(t)\right]$.

\subsection{The rumor-truth mixed spreading model}

To model the rumor-truth mixed spreading process, we need to introduce a set of hypotheses as follows.
	
\begin{enumerate}
		
\item[(H$_1$)] Due to the negative impact of rumor-believers, an uncertain person $i$ turns to believe the rumor at time $t$ at the expected rate $\beta_1 \sum_{j=1}^N a_{ji}R_j(t)$. We refer to $\beta_1 $ as the \emph{first rumor-spreading rate}.
		
\item[(H$_2$)] Due to the negative impact of rumor-believers, a truth-believer $i$ turns to believe the rumor at time $t$ at the expected rate $\beta_2 \sum_{j=1}^N a_{ji}R_j(t)$. We refer to $\beta_2$ as the \emph{second rumor-spreading rate}.		
		
\item[(H$_3$)] Due to the positive influence of truth-believers, an uncertain person $i$ turns to believe the truth at time $t$ at the expected rate $\gamma_1\sum_{j=1}^N b_{ji}T_j(t)$. We refer to $\gamma_1$ as the \emph{first truth-spreading rate}.
		
\item[(H$_4$)] Due to the positive influence of truth-believers, a rumor-believer $i$ turns to believe the truth at time $t$ at the expected rate $\gamma_2\sum_{j=1}^N b_{ji}T_j(t)$. We refer to $\gamma_2$ as the \emph{second truth-spreading rate}.
		
\item[(H$_5$)] Due to the limited memory, each rumor-believer forgets the rumor at any time at the expected rate $\delta$ and becomes uncertain, and each truth-believer forgets the truth at any time at the same expected rate $\delta$ and becomes uncertain. We refer to $\delta$ as the \emph{forgetting rate}.
		
\end{enumerate}
	
Fig. 1 schematically shows these hypotheses.
	
\begin{figure}[H]
\centering
\includegraphics[width=0.5\linewidth]{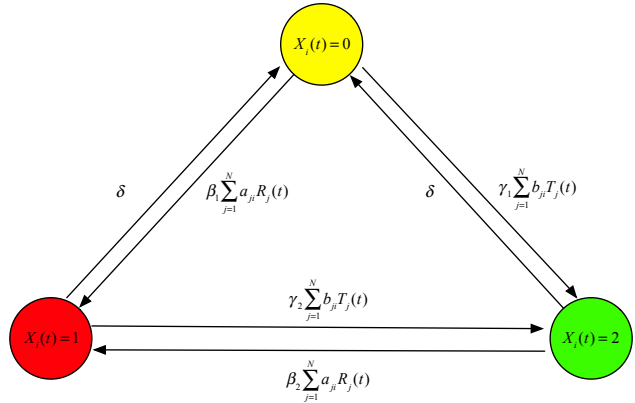}
\caption{Diagram of the hypotheses (H$_1$)--(H$_5$).}
\end{figure}
	
Based on the above hypotheses, the time evolution of the expected state of the population obeys the following differential dynamical system:
	\begin{equation}
	\left \{
	\begin{aligned}
	\frac{dR_i(t)}{dt}& = \beta_1[1-R_i(t)-T_i(t)]\sum_{j=1}^{N} a_{ji}R_j(t) + \beta_2T_i(t)\sum_{j=1}^{N} a_{ji}R_j(t)- \gamma_2R_i(t)\sum_{j=1}^{N} b_{ji}T_j(t) - \delta R_i(t), \\
	\frac{dT_i(t)}{dt}& = \gamma_1[1-R_i(t)-T_i(t)]\sum_{j=1}^{N} b_{ji}T_j(t) + \gamma_2R_i(t) \sum_{j=1}^{N} b_{ji}T_j(t) - \beta_2T_i(t)\sum_{j=1}^{N} a_{ji}R_j(t) -\delta T_i(t), \\
	& \quad \quad 0 \leq t \leq T, i = 1, 2, \cdots, N,
	\end{aligned}
	\right .
	\end{equation}
with the initial condition $\mathbf{E}(0) = \mathbf{E}^*$. We refer to the dynamical system as the \emph{uncertain-rumor-truth-uncertain (URTU) model}. A URTU model is determined by the 9-tuple 
\begin{equation}
  \mathscr{M}_{URTU} = (G_R, G_T, \beta_1, \beta_2, \gamma_1, \gamma_2, \delta, T, \mathbf{E}^*).
\end{equation}

\section{The modeling of the rumor-containing problem}
	
This section aims to model the rumor-containing (RC) problem presented at the beginning of the last section. 

From the perspective of the designer of rumor-containing strategies, the two rumor-spreading rates, $\beta_1$ and $\beta_2$, and the forgetting rate, $\delta$, are uncontrollable, but the two truth-spreading rates, $\gamma_1$ and $\gamma_2$, are controllable. Specifically, the vector $\gamma = (\gamma_1, \gamma_2)$ represents a rumor-containing strategy. For the modeling purpose, we have to introduce three additional hypotheses as follows. 

\begin{enumerate}

\item[(H$_6$)] The cost per unit time required for achieving the first truth-spreading rate $\gamma_1$ is $c_1 \gamma_1$ units. We refer to $c_1$ as the \emph{first truth-spreading cost coefficient}.

\item[(H$_7$)] The cost per unit time required for achieving the second truth-spreading rate $\gamma_2$ is $c_2 \gamma_2$ units. We refer to $c_2$ as the \emph{second truth-spreading cost coefficient}.

\item[(H$_8$)] The budget per unit time used for containing the rumor is $B$, i.e., $c_1 \gamma_1 + c_2 \gamma_2 = B$. 

\end{enumerate}

Given a rumor-containing strategy $\gamma = (\gamma_1, \gamma_2)$. It is seen from the URTU model (4) that the expected cumulative number of uncertain persons who turn to believe the truth is 
\begin{equation}
  E_U(\gamma) = \gamma_1\int_0^T\sum_{i=1}^N[1 - R_i(t) - T_i(t)]\sum_{j=1}^{N} b_{ji}T_j(t)dt,
\end{equation}
and the expected cumulative number of rumor-believers who turn to believe the truth is
\begin{equation}
  E_R(\gamma) = \gamma_2\int_0^T\sum_{i=1}^NR_i(t)\sum_{j=1}^{N} b_{ji}T_j(t)dt
\end{equation}
Hence, the expected cumulative number of uncertain persons or rumor-believers who turn to believe the truth is
\begin{equation}
\begin{split}
	E(\gamma) &= E_U(\gamma) + E_R(\gamma) \\
	&= \gamma_1\int_0^T\sum_{i=1}^N[1 - R_i(t) - T_i(t)]\sum_{j=1}^{N} b_{ji}T_j(t)dt + \gamma_2\int_0^T\sum_{i=1}^NR_i(t)\sum_{j=1}^{N} b_{ji}T_j(t)dt
\end{split}
\end{equation}
Naturally, we are going to take $E(\gamma)$ as the measure of the \emph{effectiveness} of the rumor-containing strategy $\gamma$.

Let $\Omega$ denote the admissible set of rumor-containing strategies. In view of hypothesis (H$_8$), we have
\begin{equation}
 \Omega = \left\{\gamma = (\gamma_1, \gamma_2) \in \mathbb{R}_+^2 : c_1 \gamma_1 + c_2 \gamma_2 = B\right\}.
\end{equation}
Therefore, we are going to take the quantity
\begin{equation}
\frac{E(\gamma)}{BT} = \frac{1}{BT}\left[\gamma_1\int_0^T\sum_{i=1}^N[1 - R_i(t) - T_i(t)]\sum_{j=1}^{N} b_{ji}T_j(t)dt + \gamma_2\int_0^T\sum_{i=1}^NR_i(t)\sum_{j=1}^{N} b_{ji}T_j(t)dt\right]
\end{equation} 
as the measure of the \emph{cost effectiveness} of the rumor-containing strategy  $\gamma$.

Based on the above discussions, the original RC problem is modeled as the following constrained optimization problem: 
\begin{equation}
\begin{split}
  \max_{\gamma \in \Omega} E(\gamma) &= \gamma_1\int_0^T\sum_{i=1}^N[1 - R_i(t) - T_i(t)]\sum_{j=1}^{N} b_{ji}T_j(t)dt + \gamma_2\int_0^T\sum_{i=1}^NR_i(t)\sum_{j=1}^{N} b_{ji}T_j(t)dt\\
  & \text{ s.t. URTU model (6)}.
\end{split}  
\end{equation}
Here, an optimal solution stands for a most effective rumor-containing strategy, and the optimal value stands for the maximum possible effectiveness over all admissible rumor-containing strategies. We refer to the optimization problem as the \emph{rumor-containing (RC) model}. A RC model is determined by the 10-tuple 
\begin{equation}
  \mathscr{M}_{RC} = (G_R, G_T, \beta_1, \beta_2, \delta, T, B, c_1, c_2, \mathbf{E}^*).
\end{equation}

The RC model can be written in reduced form as
\begin{equation}
\begin{split}
\max_{0 \leq \gamma_1 \leq \frac{B}{c_1}} \hat{E}(\gamma_1) &= \gamma_1\int_0^T\sum_{i=1}^N[1 - R_i(t) - T_i(t)]\sum_{j=1}^{N} b_{ji}T_j(t)dt + \frac{B - c_1 \gamma_1}{c_2}\int_0^T\sum_{i=1}^NR_i(t)\sum_{j=1}^{N} b_{ji}T_j(t)dt\\
& \text{ s.t. URTU model (6)}.
\end{split}  
\end{equation}

In practice, the most effective rumor-containing strategies are recommended, and the two truth-spreading rates in a most effective rumor-containing strategy can be enhanced by improving people's education level or/and formulating a perfect policy of restraining false rumors.

\section{Some most effective rumor-containing strategies}

In the last section, the RC model characterizing the RC problem was established. The next thing to do is to find the most effective rumor-containing strategy by solve a RC model. Due to the inherent complexity of the RC model, it seems impracticable to cope with the model analytically. In this section, let us give some most effective rumor-containing strategies by solving the respective RC models. 

Fig. 2 plots all the connected graphs with two to four nodes, up to isomorphism.

\begin{figure}[H]
	\centering
	\subfigure[$G_1$]{\includegraphics[width=0.15\textwidth]{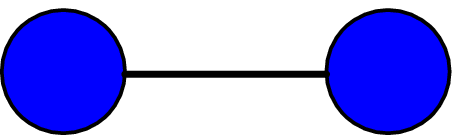}}
	\hspace{.2in}
	\subfigure[$G_2$]{\includegraphics[width=0.15\textwidth]{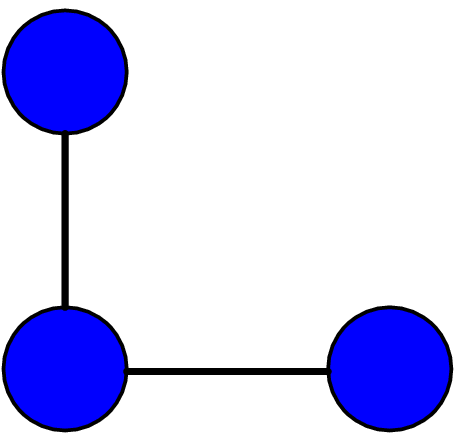}}
	\hspace{.2in}
	\subfigure[$G_3$]{\includegraphics[width=0.15\textwidth]{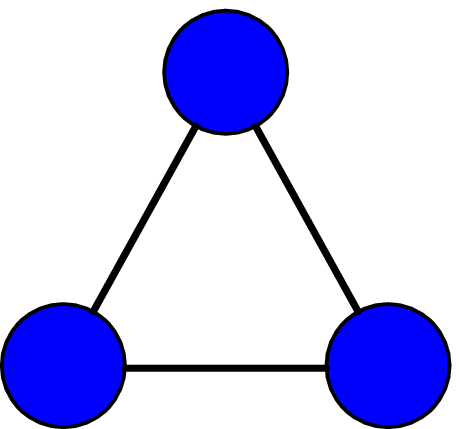}}
	\\
	\subfigure[$G_4$]{\includegraphics[width=0.20\textwidth]{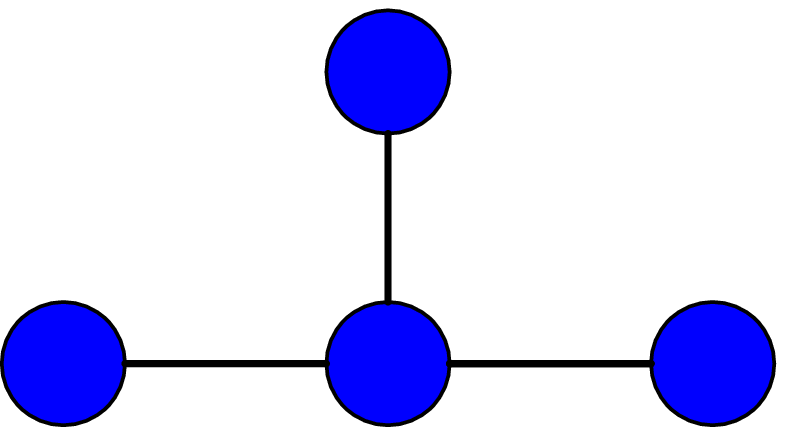}}
	\hspace{.2in}
	\subfigure[$G_5$]{\includegraphics[width=0.15\textwidth]{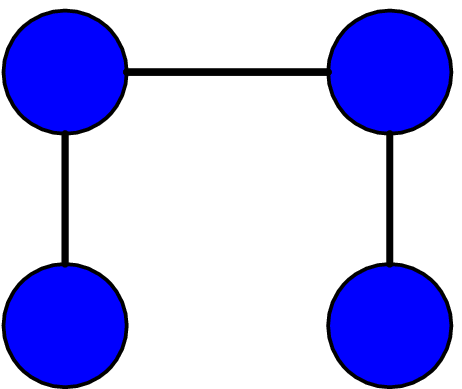}}
	\hspace{.2in}
	\subfigure[$G_6$]{\includegraphics[width=0.15\textwidth]{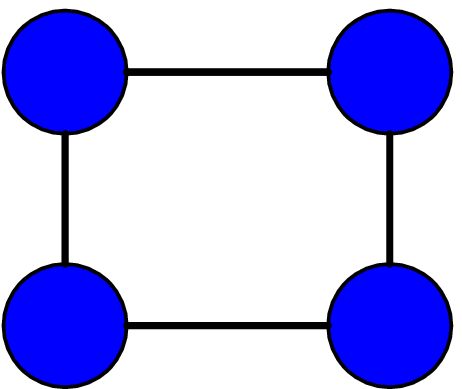}}
	\\
	\subfigure[$G_7$]{\includegraphics[width=0.15\textwidth]{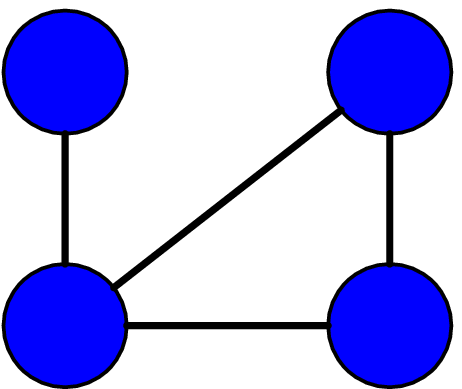}}
	\hspace{.2in}
	\subfigure[$G_8$]{\includegraphics[width=0.15\textwidth]{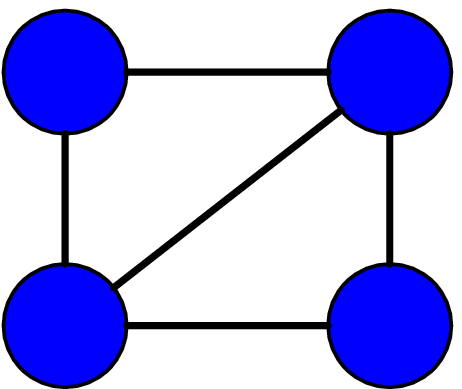}}
	\hspace{.2in}
	\subfigure[$G_9$]{\includegraphics[width=0.15\textwidth]{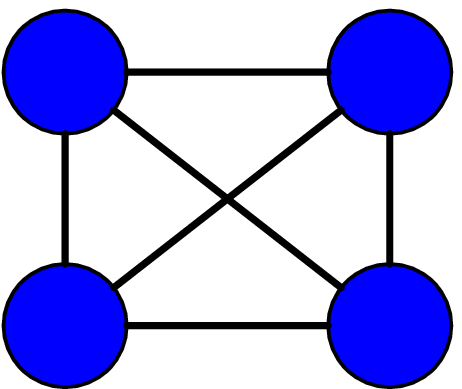}}
	\caption{Up to isomorphism, all connected graphs with two to four nodes.} 
	\label{fig:subfig} 
\end{figure}

\begin{expe}
Consider the following RC models:
\begin{tabbing}
  \hspace{1.5cm} $\mathscr{M}_{RC}$ \quad\quad\= $G_R$ \quad\quad\= $G_T$ \quad\quad\= $\beta_1$ \quad\quad\= $\beta_2$ \quad\quad\= $\delta$ \quad\quad\= T \quad\quad\= c$_1$ \quad\quad\= c$_2$ \quad\quad\= B \quad\quad\= $\mathbf{E}^*$\\
  \hspace{1.5cm} $\mathscr{M}_{RC}^{(1)}$ \quad\quad\= $G_1$ \> $G_1$ \> 0.7 \> 0.1 \> 0.1 \> 35 \> 8 \>3 \>10 \> $[0.1, \cdots, 0.1]$ \\
  \hspace{1.5cm} $\mathscr{M}_{RC}^{(2)}$ \quad\quad\= $G_2$ \> $G_3$ \> 0.7 \> 0.6 \> 0.7 \> 55 \> 4 \>8 \>6 \> $[0.1, \cdots, 0.1]$
  \\
  \hspace{1.5cm} $\mathscr{M}_{RC}^{(3)}$ \quad\quad\= $G_3$ \> $G_2$ \> 0.4 \> 0.3 \> 0.3 \> 35 \> 4 \>6 \> 6 \> $[0.1, \cdots, 0.1]$
  \\
  \hspace{1.5cm} $\mathscr{M}_{RC}^{(4)}$ \quad\quad\= $G_4$ \> $G_5$ \> 0.9 \> 0.7 \> 0.7 \> 50 \> 3 \>3 \>6 \> $[0.1, \cdots, 0.1]$
  \\
  \hspace{1.5cm} $\mathscr{M}_{RC}^{(5)}$ \quad\quad\= $G_5$ \> $G_6$ \> 0.1 \> 0.8 \> 0.4 \> 50 \> 5 \>5 \>6 \> $[0.1, \cdots, 0.1]$
  \\
  \hspace{1.5cm} $\mathscr{M}_{RC}^{(6)}$ \quad\quad\= $G_6$ \> $G_7$ \> 0.5 \> 0.8 \> 0.5 \> 70 \> 2 \>4 \>6 \> $[0.1, \cdots, 0.1]$
  \\
  \hspace{1.5cm} $\mathscr{M}_{RC}^{(7)}$ \quad\quad\= $G_7$ \> $G_8$ \> 0.1 \> 0.4 \> 0.2 \> 40 \> 4 \>3 \>4 \> $[0.1, \cdots, 0.1]$
  \\
  \hspace{1.5cm} $\mathscr{M}_{RC}^{(8)}$ \quad\quad\= $G_8$ \> $G_9$ \> 0.5 \> 0.5 \> 0.7 \> 70 \> 9 \>9 \>12 \> $[0.1, \cdots, 0.1]$
  \\
  \hspace{1.5cm} $\mathscr{M}_{RC}^{(9)}$ \quad\quad\= $G_9$ \> $G_4$ \> 0.9 \> 0.6 \> 0.1 \> 45 \> 3 \>3 \>6 \> $[0.1, \cdots, 0.1]$
\end{tabbing}
Fig. 3 exhibits all the optimal solutions to these RC models obtained by numerically solving the models.
\end{expe}

\begin{figure}[H]
	\centering
	\subfigure[]{\includegraphics[width=0.25\textwidth]{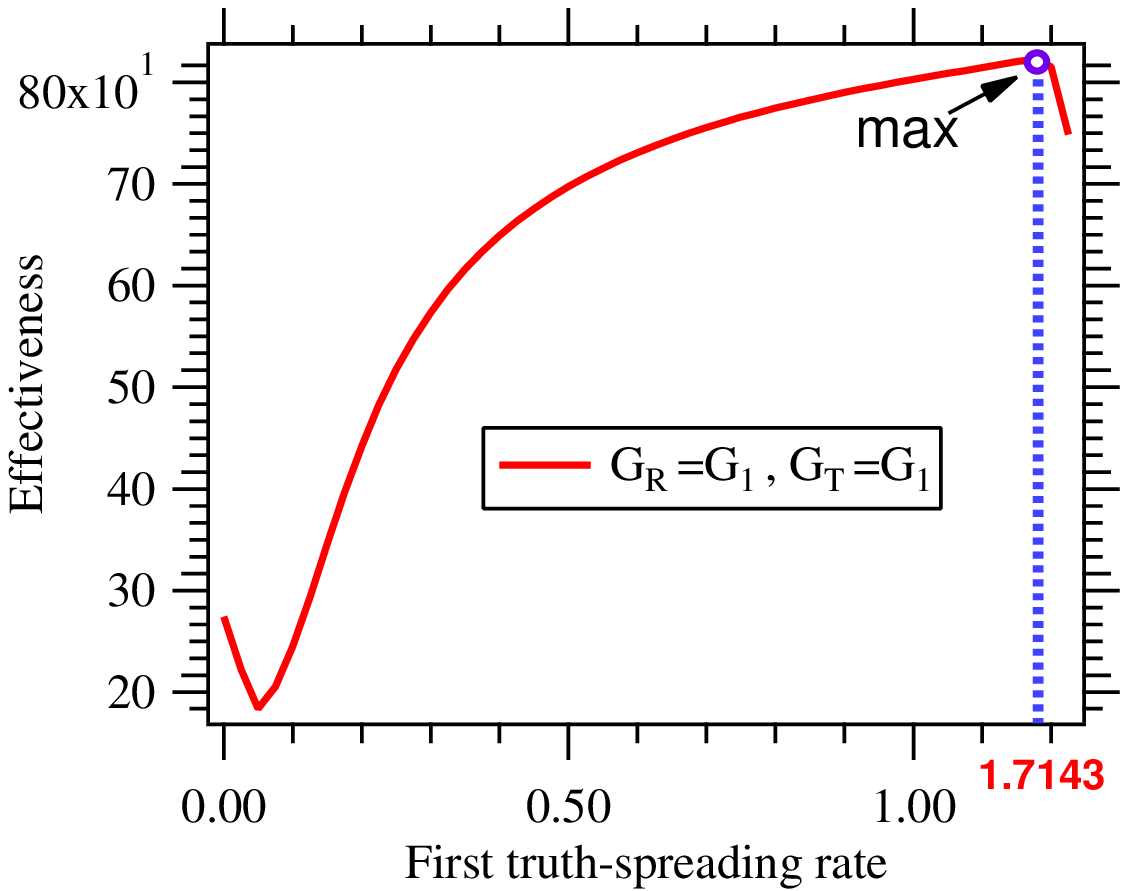}}
	\hspace{.2in}
	\subfigure[]{\includegraphics[width=0.25\textwidth]{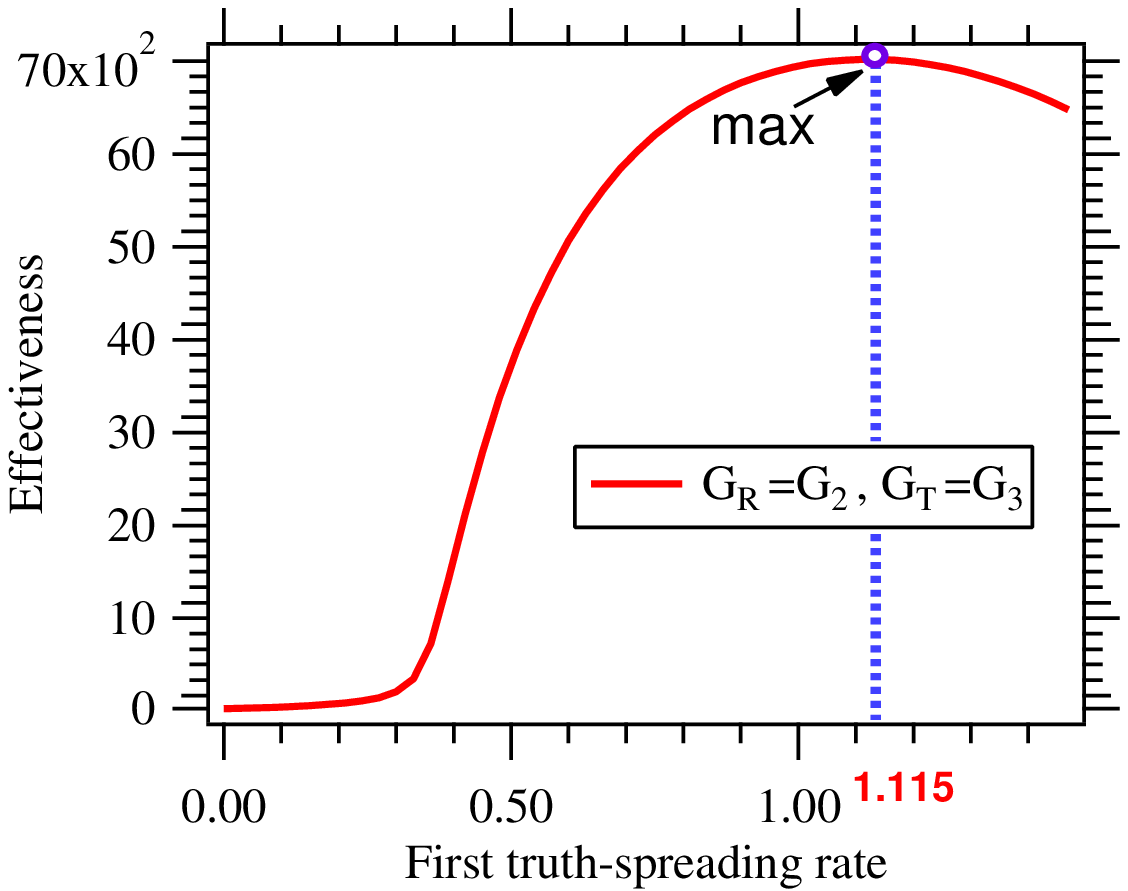}}
	\hspace{.2in}
	\subfigure[]{\includegraphics[width=0.25\textwidth]{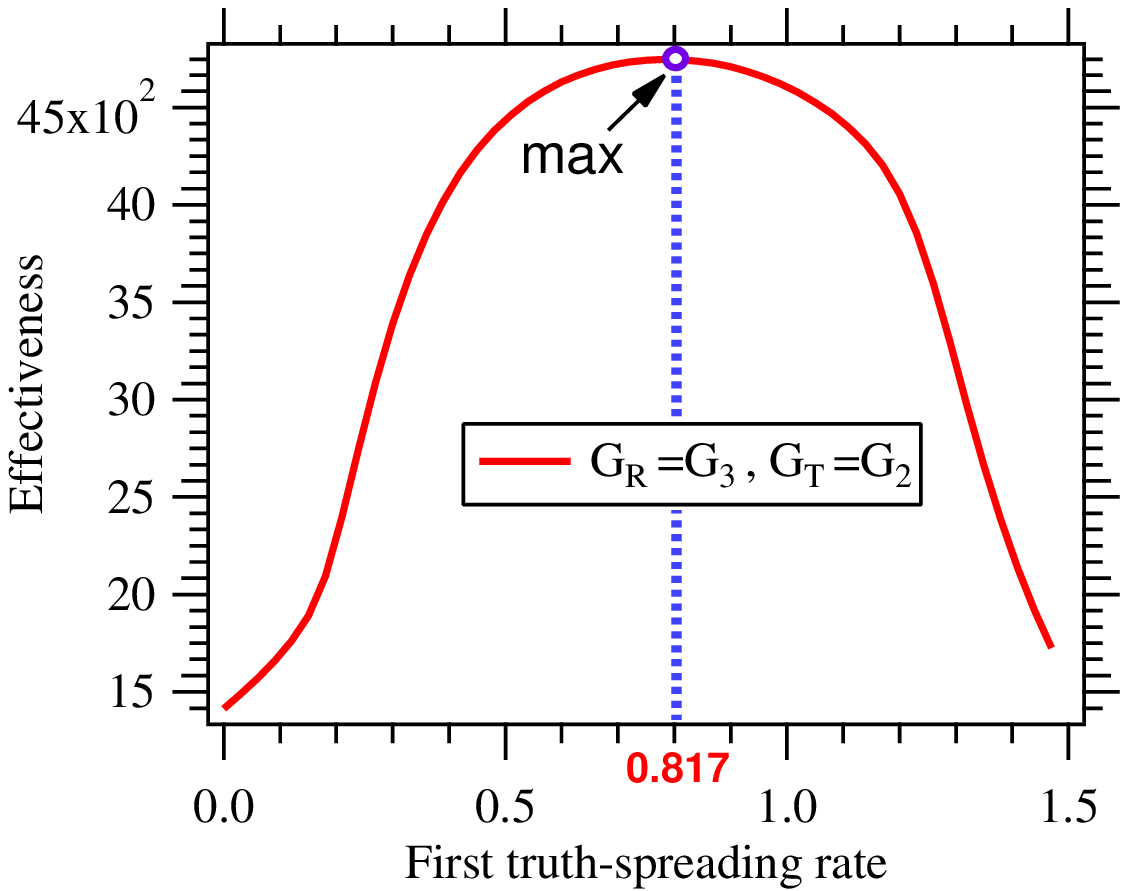}}
	\\
	\subfigure[]{\includegraphics[width=0.25\textwidth]{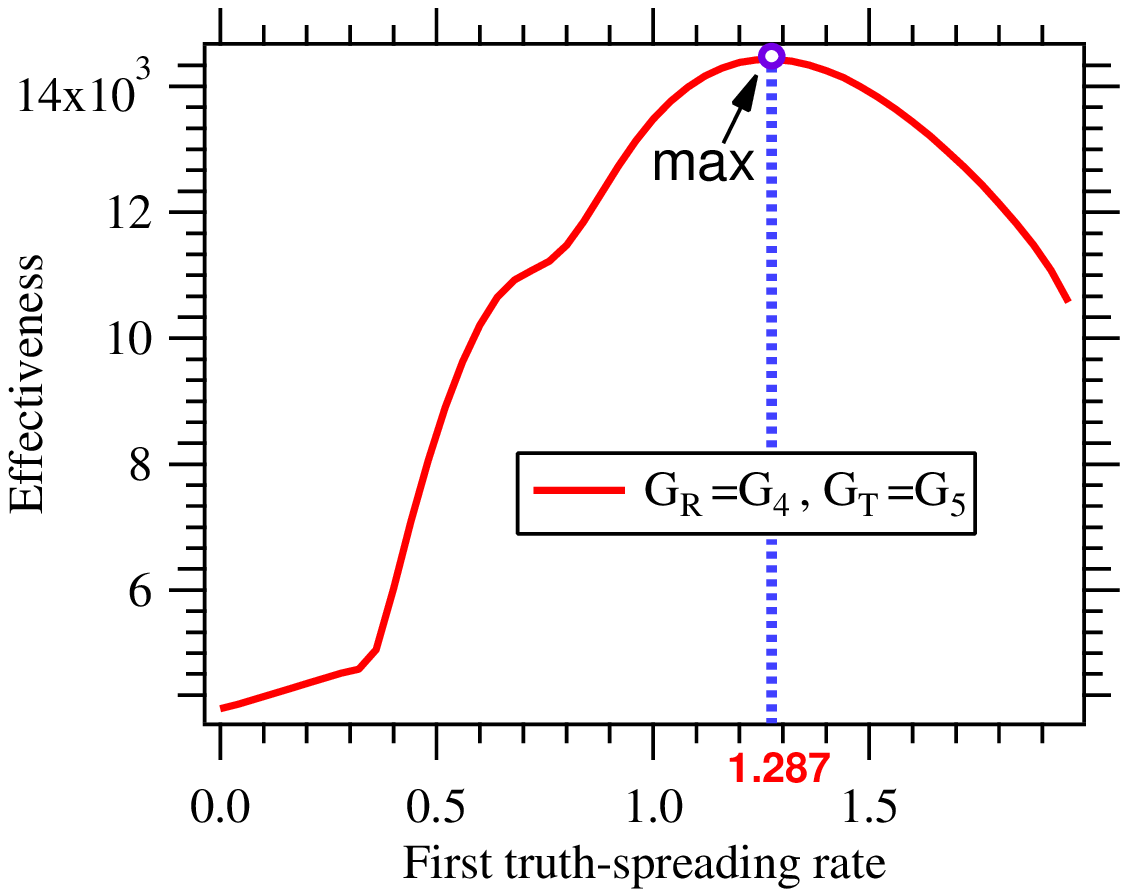}}
	\hspace{.2in}
	\subfigure[]{\includegraphics[width=0.25\textwidth]{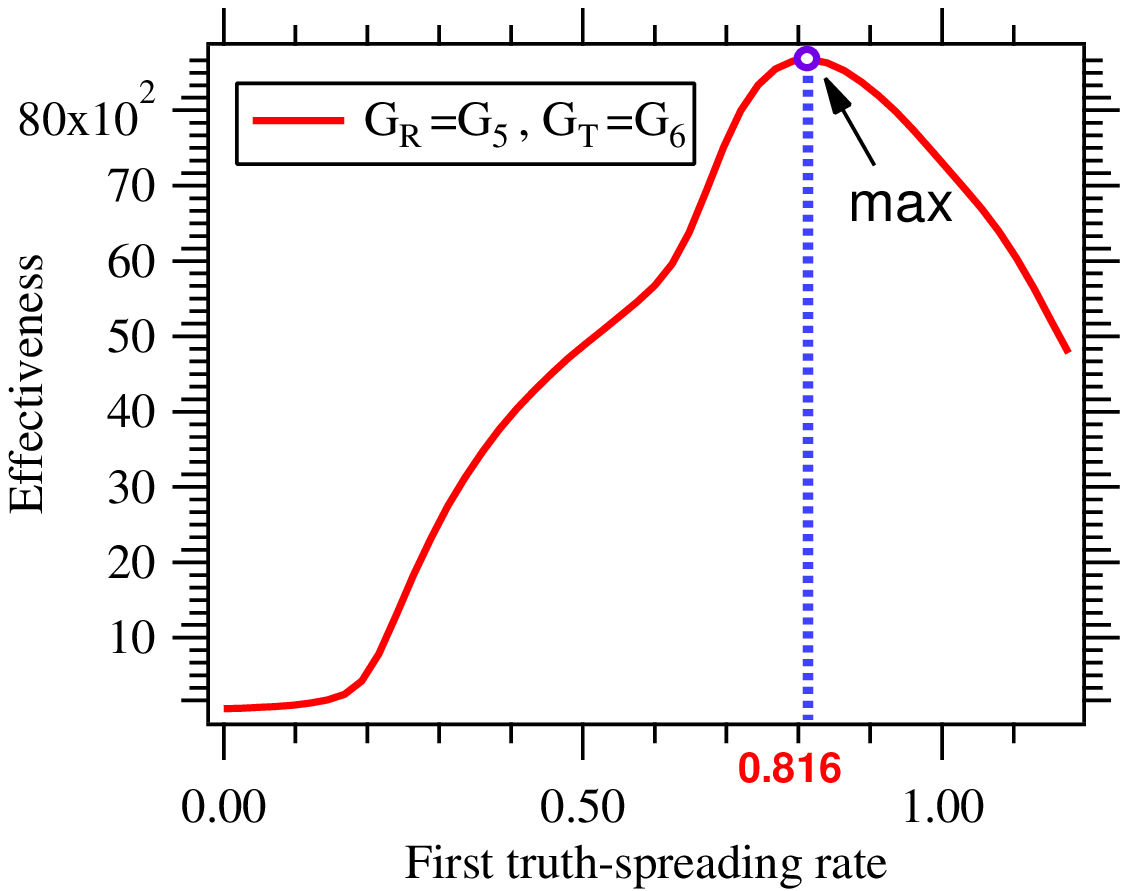}}
	\hspace{.2in}
	\subfigure[]{\includegraphics[width=0.25\textwidth]{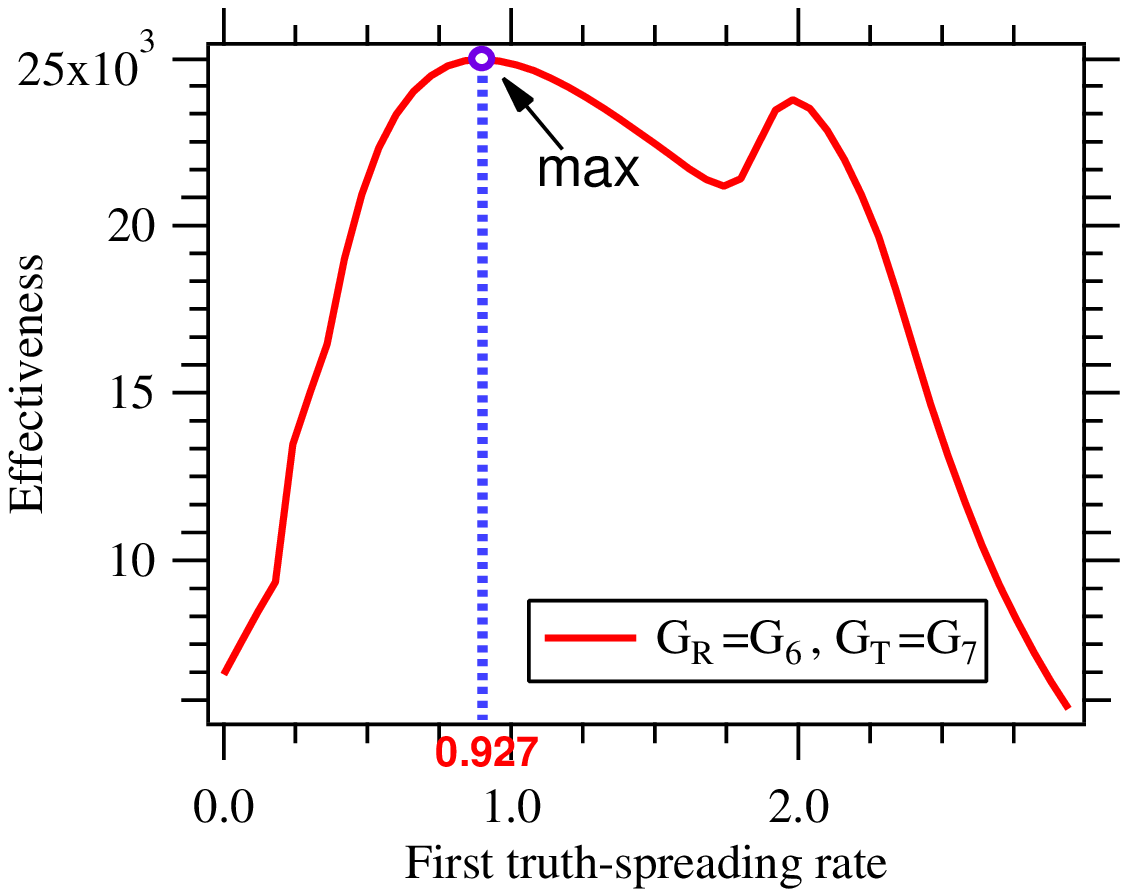}}
	\\
	\subfigure[]{\includegraphics[width=0.25\textwidth]{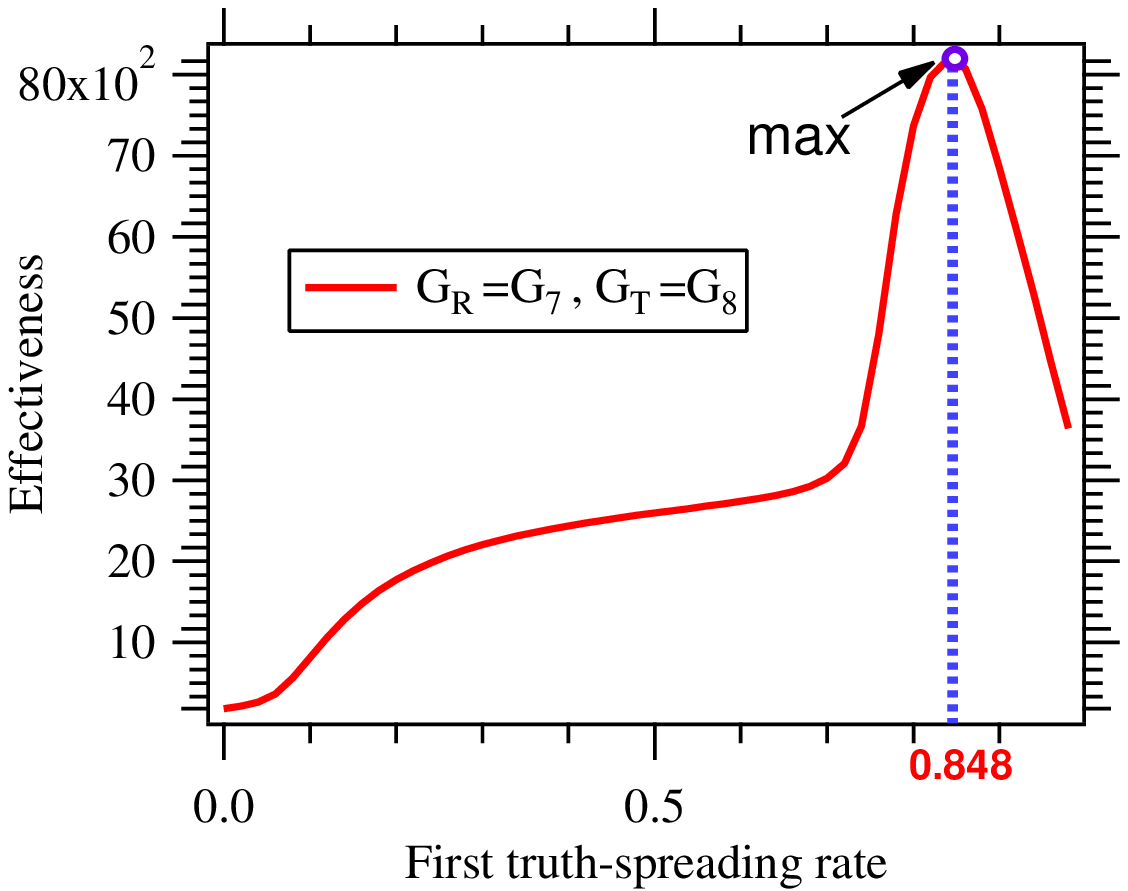}}
	\hspace{.2in}
	\subfigure[]{\includegraphics[width=0.25\textwidth]{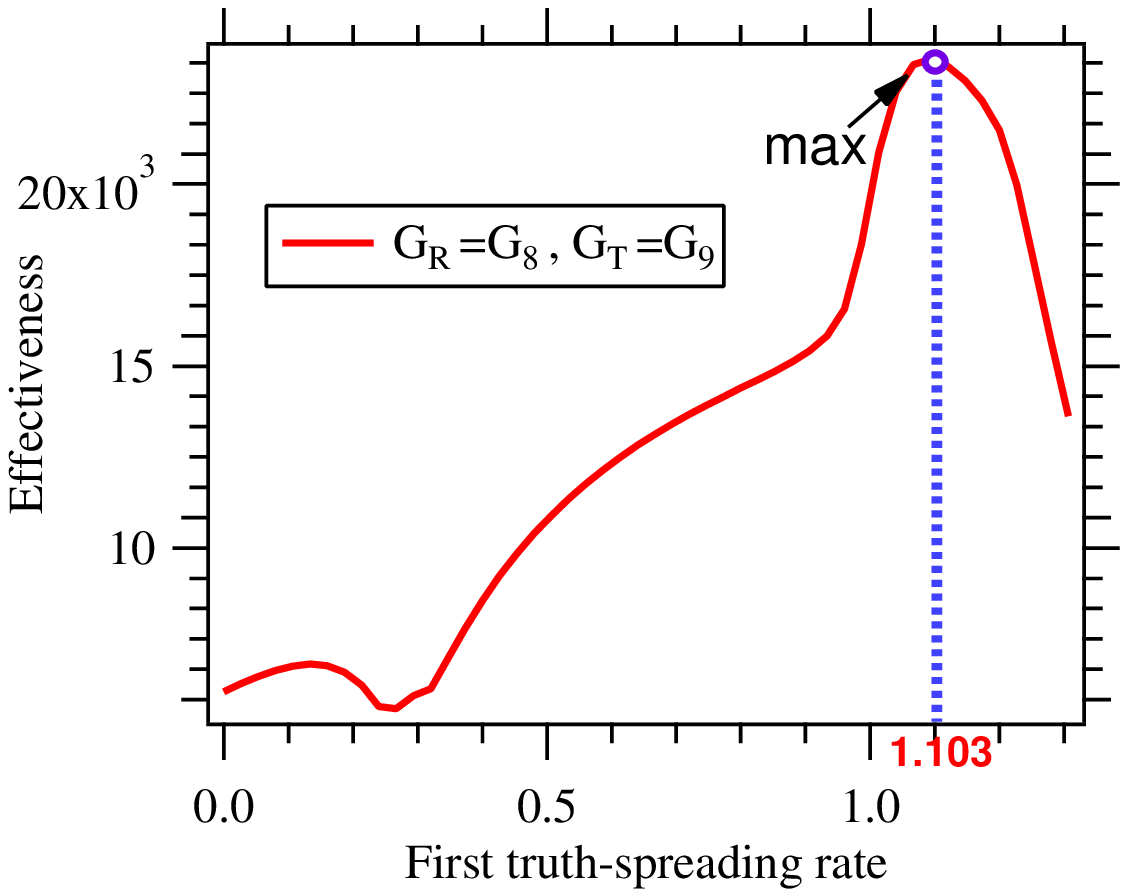}}
	\hspace{.2in}
	\subfigure[]{\includegraphics[width=0.25\textwidth]{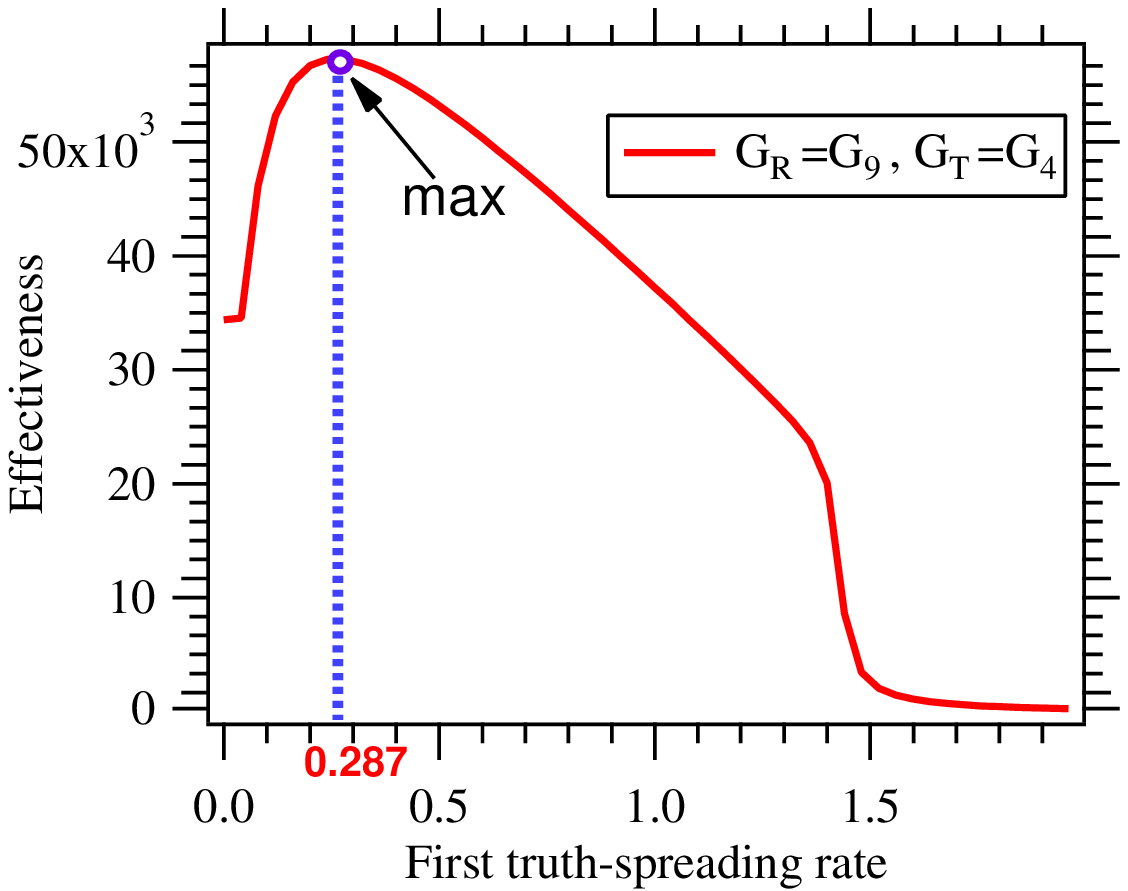}}
	\caption{The optimal solutions in Experiment 1.} 
	\label{fig:subfig} 
\end{figure}

Small-world networks are networks that are generated by randomly rewiring some edges of regular networks. Fig. 4(a) plots a small-world network with 50 nodes, which is obtained by executing the algorithm proposed by Watts and Strogatz \cite{Watts1998}. Let $G_{SW}$ denote the network. Scale-free networks are networks with an approximate power-law degree distribution. Fig. 4(b) depicts a scale-free network with 50 nodes, which is obtained by executing the algorithm proposed by Barabasi and Albert \cite{Barabasi1999}. Let $G_{SF}$ denote the network. Fig. 4(c) exhibits a realistic network with 49 nodes, which comes from Ref. \cite{konect}. Let $G_{RE}$ denote the network.

\begin{figure}[H]
	\centering
	\subfigure[The small-world network $G_{SW}$.]{\includegraphics[width=0.3\textwidth]{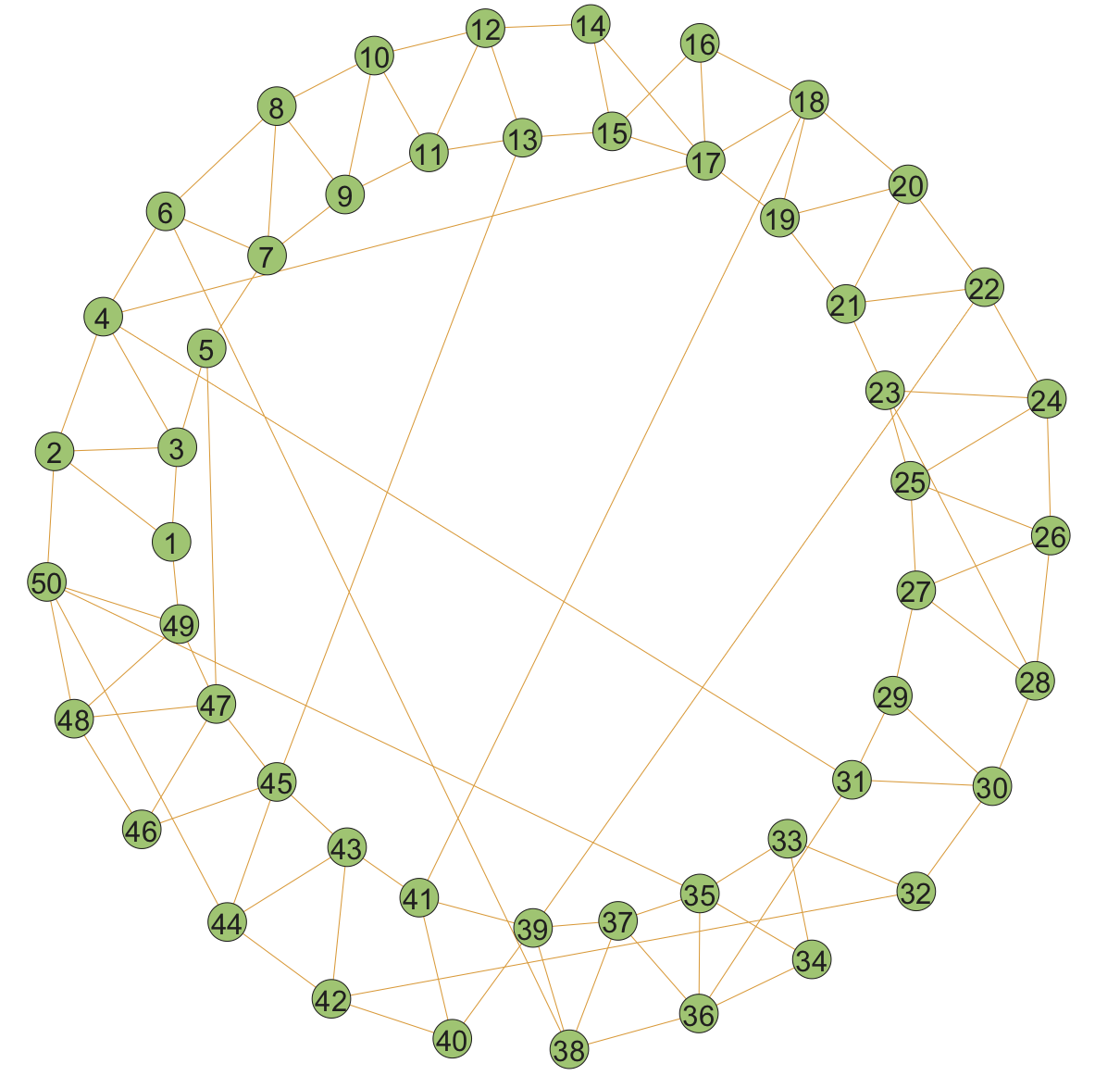}}
	\hspace{.2in}
	\subfigure[The scale-free network $G_{SF}$.]{\includegraphics[width=0.3\textwidth]{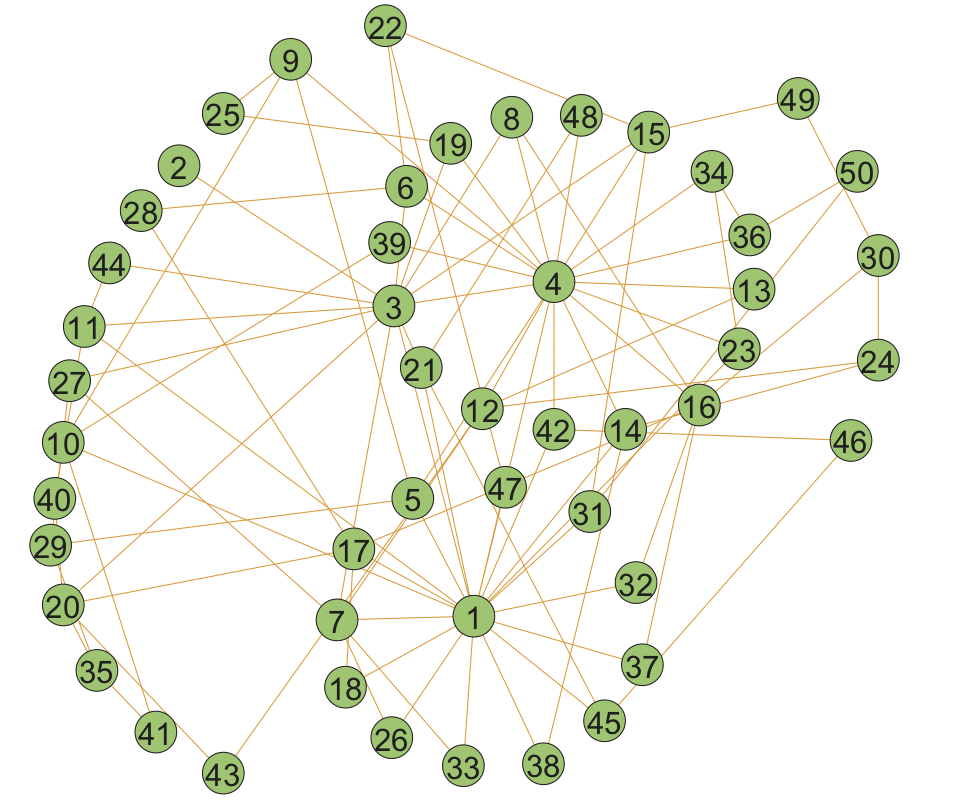}}
	\hspace{.2in}
	\subfigure[The realistic network $G_{RE}$.]{\includegraphics[width=0.3\textwidth]{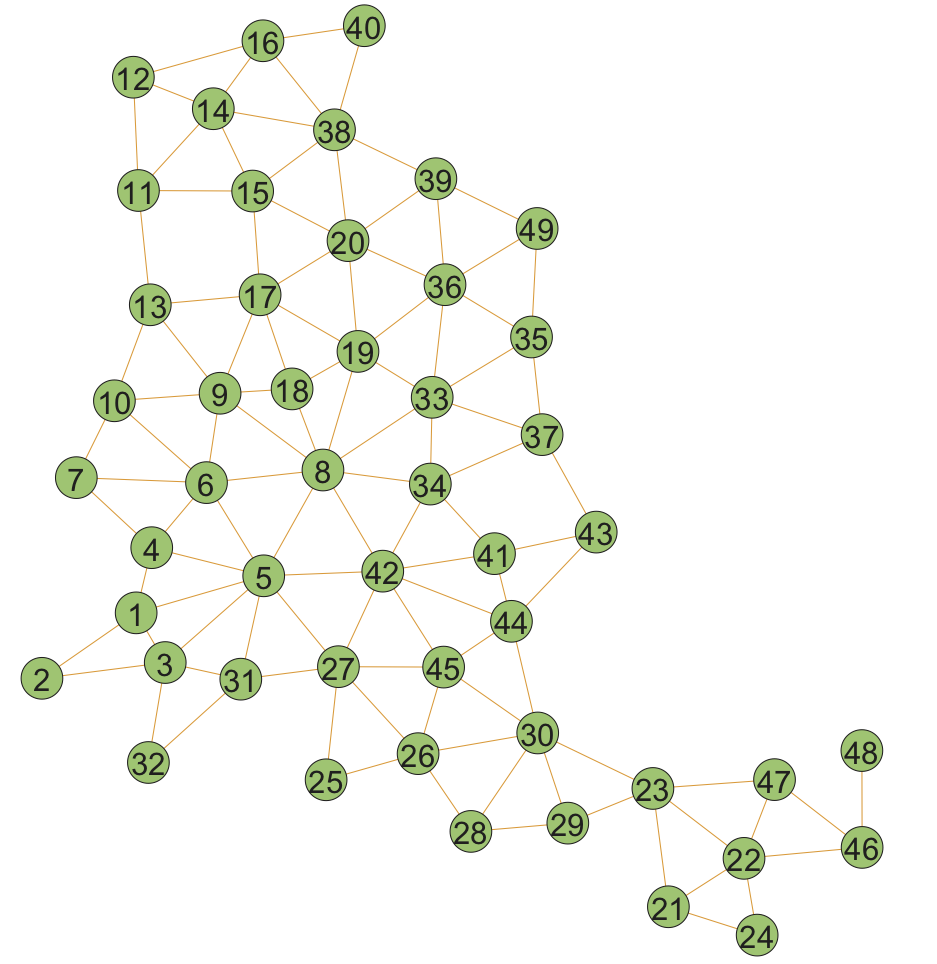}}
	\centering
	\caption{Three networks.}
\end{figure}

\begin{expe}
Consider the following RC models:
\begin{tabbing}
  \hspace{1.5cm} $\mathscr{M}_{RC}$ \quad\quad\= $G_R$ \quad\quad\= $G_T$ \quad\quad\= $\beta_1$ \quad\quad\= $\beta_2$ \quad\quad\= $\delta$ \quad\quad\= T \quad\quad\= $c_1$ \quad\quad\= $c_2$ \quad\quad\= B \quad\quad\= $\mathbf{E}^*$\\
  \hspace{1.5cm} $\mathscr{M}_{RC}^{(1)}$ \quad\quad\= $G_{SW}$ \> $G_{SF}$ \> 0.4 \> 0.7 \> 0.5 \> 30 \> 3 \>9 \>18 \> $[0.1, \cdots, 0.1]$\\
  \hspace{1.5cm} $\mathscr{M}_{RC}^{(2)}$ \quad\quad\= $G_{SF}$ \> $G_{SW}$ \> 0.6 \> 0.8 \> 0.2 \> 50 \> 2 \>2 \>2 \> $[0.1, \cdots, 0.1]$\\
  \hspace{1.5cm} $\mathscr{M}_{RC}^{(3)}$ \quad\quad\= $G_{RE}$ \> $G_{RE}$ \> 0.3 \> 0.4 \> 0.4 \> 70 \> 5 \>6 \>2 \> $[0.1, \cdots, 0.1]$
\end{tabbing}
Fig. 5 exhibits all the optimal solutions to these RC models obtained by numerically solving the models.
\end{expe}

\begin{figure}[H]
	\centering
	\subfigure[]{\includegraphics[width=0.3\textwidth]{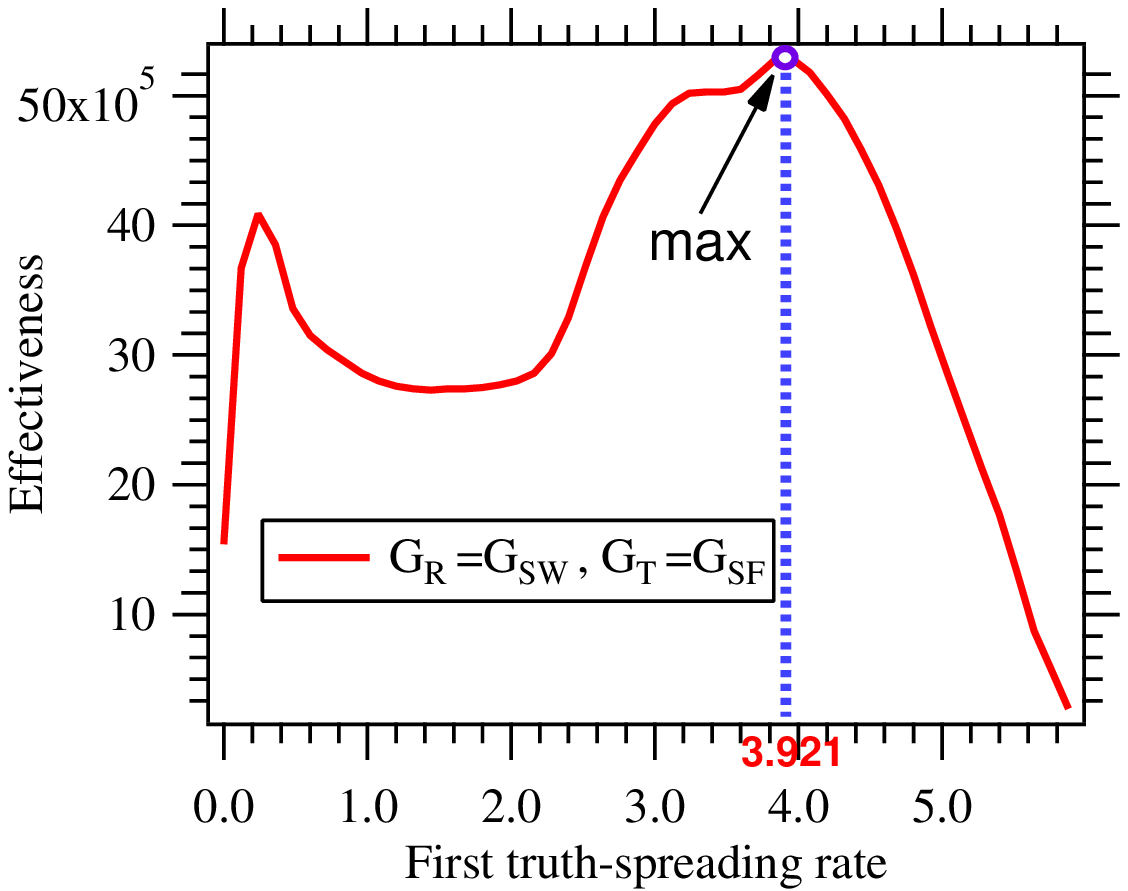}}
	\hspace{.2in}
	\subfigure[]{\includegraphics[width=0.3\textwidth]{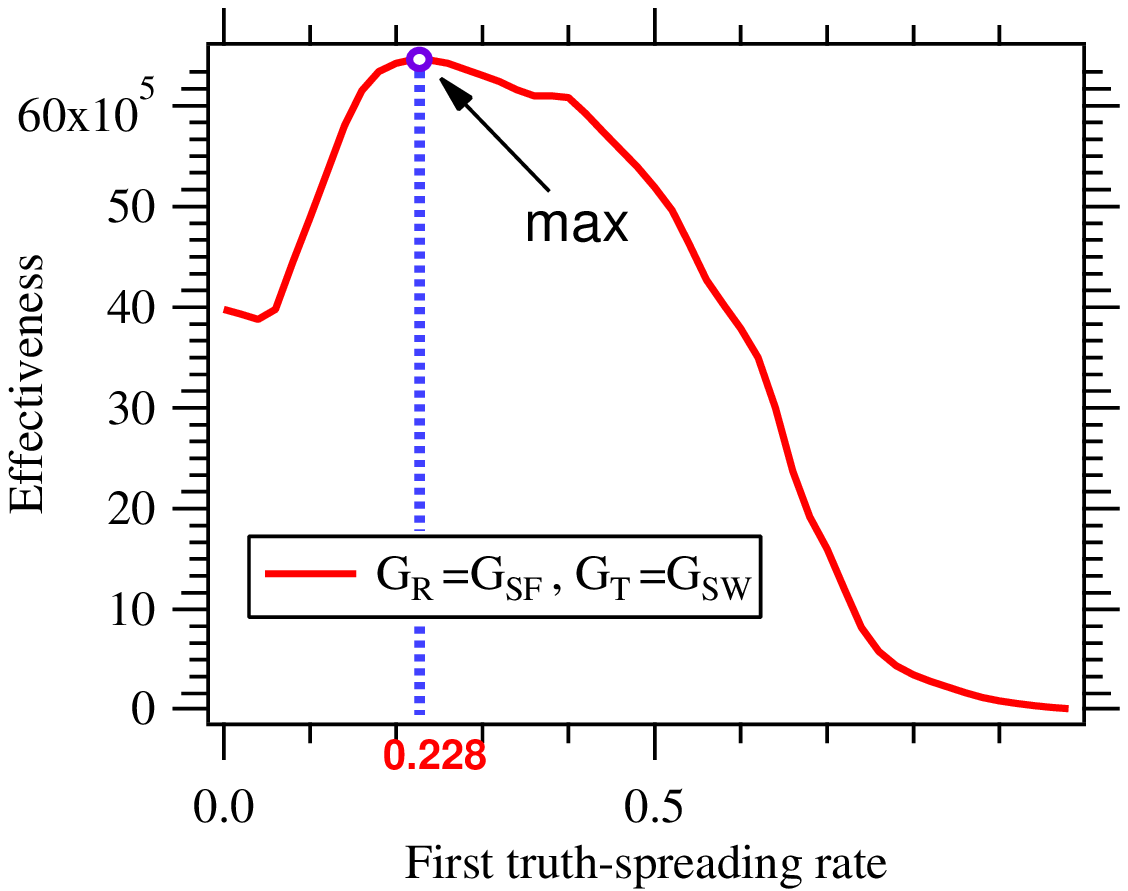}}
	\hspace{.2in}
	\subfigure[]{\includegraphics[width=0.3\textwidth]{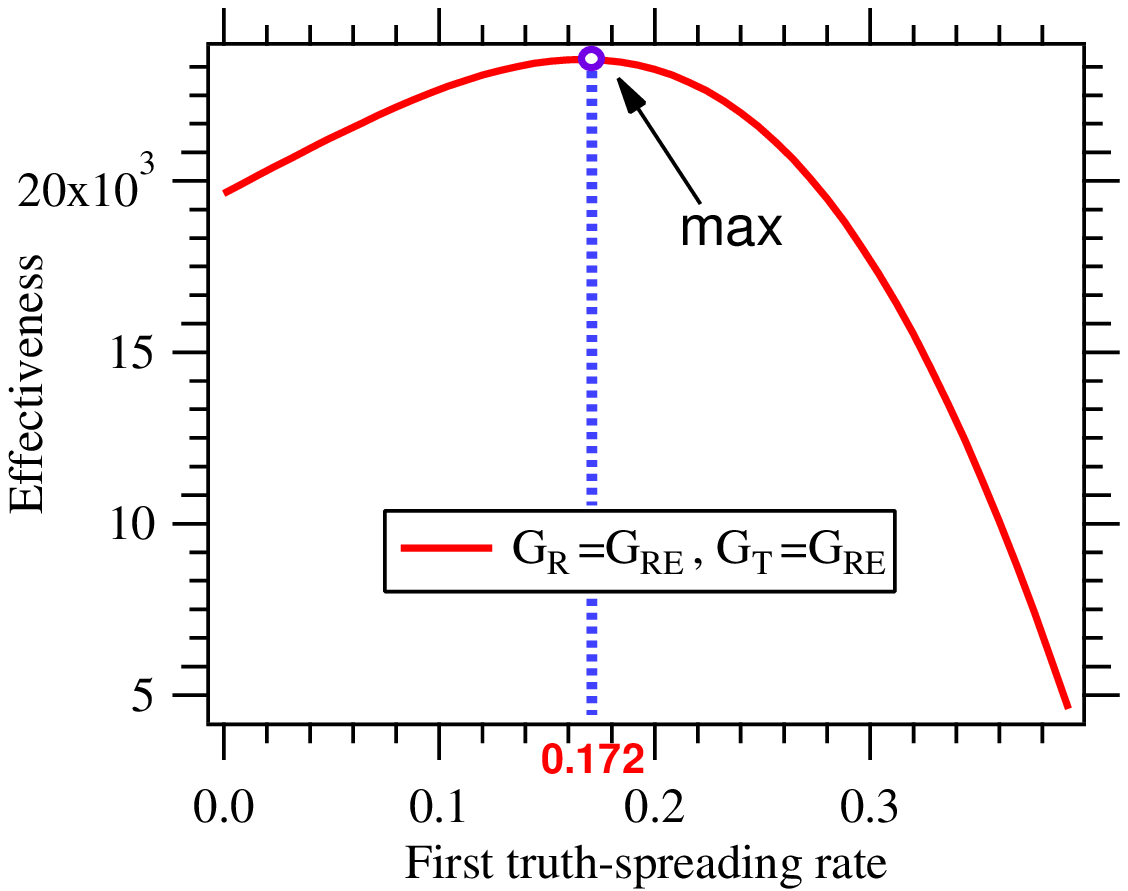}}
	\caption{The optimal solutions in Experiment 2.} 
	\label{fig:subfig} 
\end{figure}

\section{Further discussions}
	
Obviously, the highest cost effectiveness of a RC model depends on all the parameters in the RC model. The goal of this section is to uncover the way that these factors affect the highest cost effectiveness through computer experiments.

\subsection{The influence of the two rumor-spreading rates} 
	
First, let us examine the influence of the two rumor-spreading rates on the highest cost effectiveness. 

\begin{expe}
Consider the following RC models:
\begin{tabbing}
	\hspace{1.5cm} $G_R$ \quad\quad\= $G_T$ \quad\quad\= $\beta_1$ \quad\quad\quad\quad\quad\quad\= $\beta_2$ \quad\quad\= $\delta$ \quad\quad\= T \quad\quad\= B \quad\quad\= c$_1$ \quad\quad\= c$_2$ \quad\quad\= $\mathbf{E}^*$\\
	\hspace{1.5cm} $G_{SW}$ \> $G_{SF}$ \> 0.1/0.2/$\cdots$/0.9 \> 0.1 \> 0.3 \> 10 \> 6 \>1 \>2 \> $[0.1, \cdots, 0.1]$\\
	\hspace{1.5cm} $G_{SF}$ \> $G_{SW}$ \> 0.1/0.2/$\cdots$/0.9 \> 0.2 \> 0.2 \> 15 \> 8 \>2 \>3 \> $[0.1, \cdots, 0.1]$\\
	\hspace{1.5cm} $G_{RE}$ \> $G_{RE}$ \> 0.1/0.2/$\cdots$/0.9 \> 0.3 \> 0.1 \> 20 \> 10 \>3 \>4 \> $[0.1, \cdots, 0.1]$
\end{tabbing}
Fig. 6 plots the highest cost effectivenesses for these RC models. It is seen that the highest cost effectiveness is increasing with $\beta_1$. 
\end{expe}
	
\begin{figure}[H]
	\centering
	\subfigure[]{\includegraphics[width=0.3\textwidth]{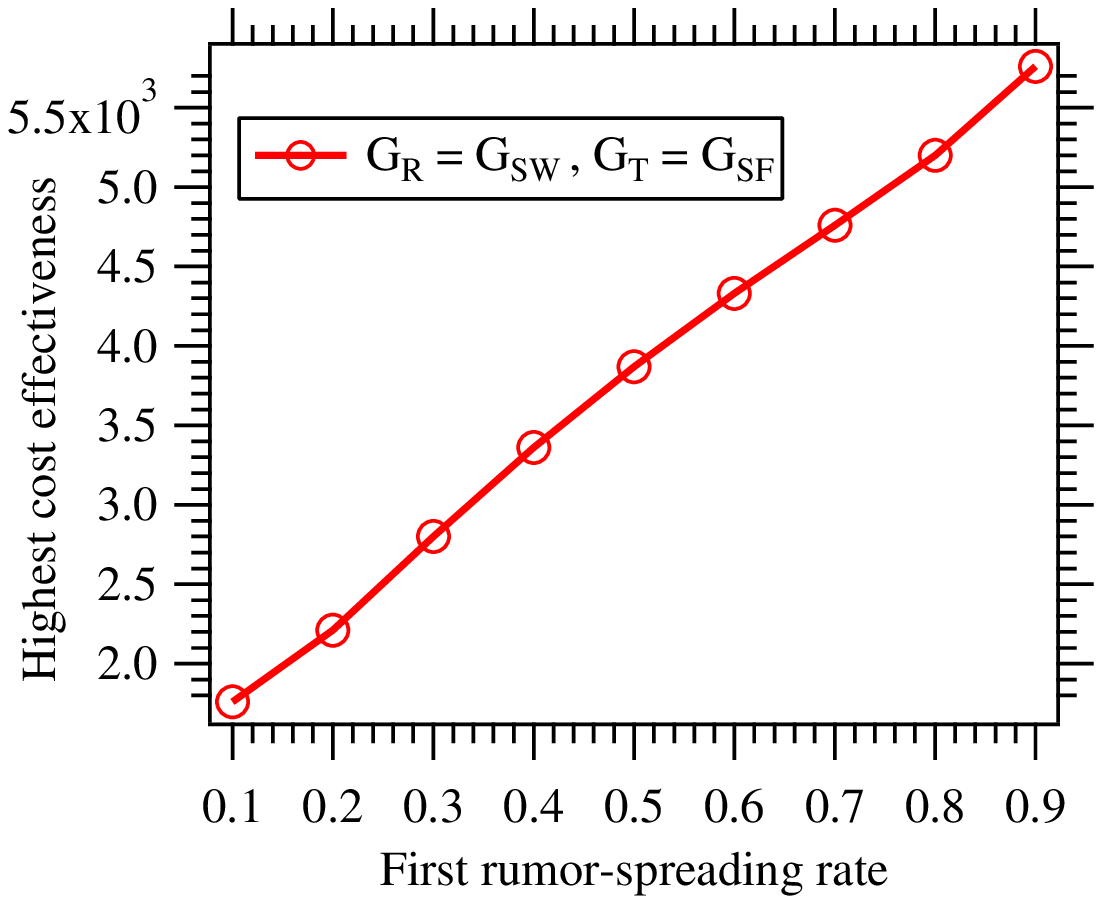}}
	\hspace{.2in}
	\subfigure[]{\includegraphics[width=0.3\textwidth]{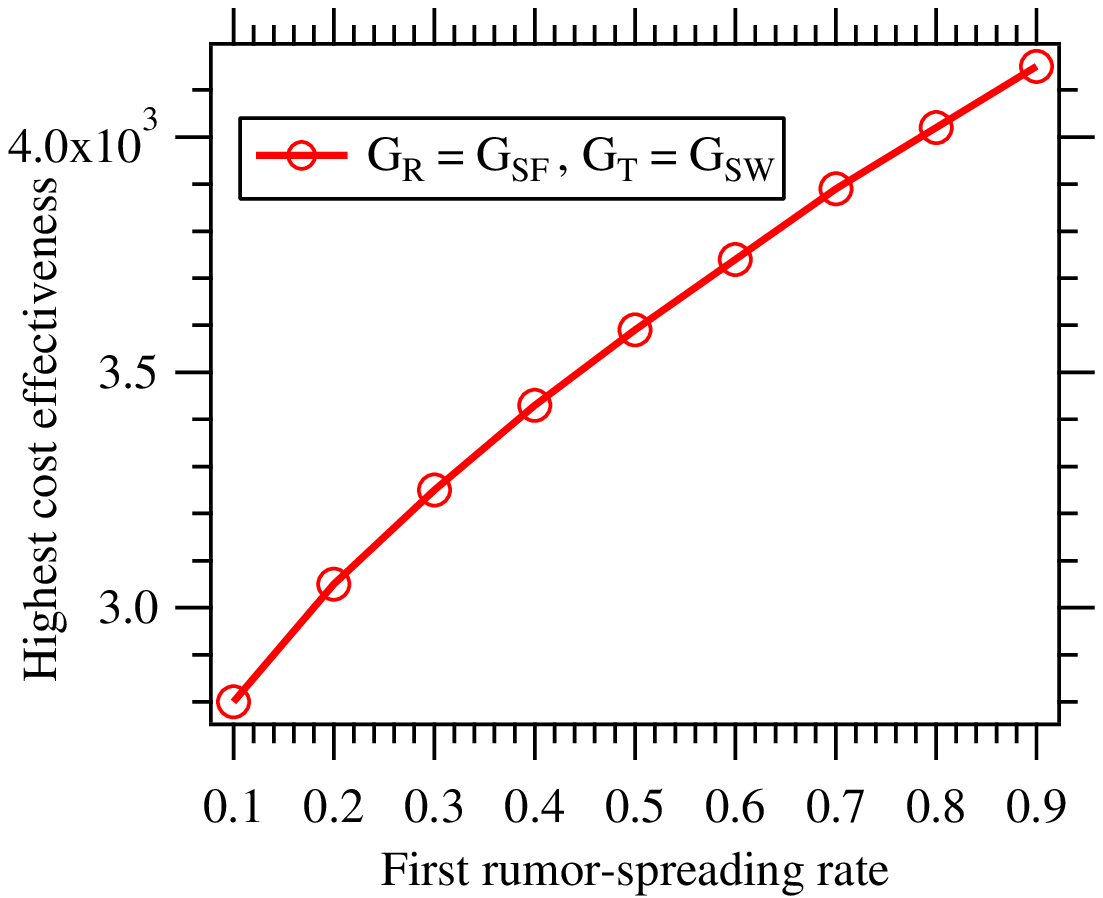}}
	\hspace{.2in}
	\subfigure[]{\includegraphics[width=0.3\textwidth]{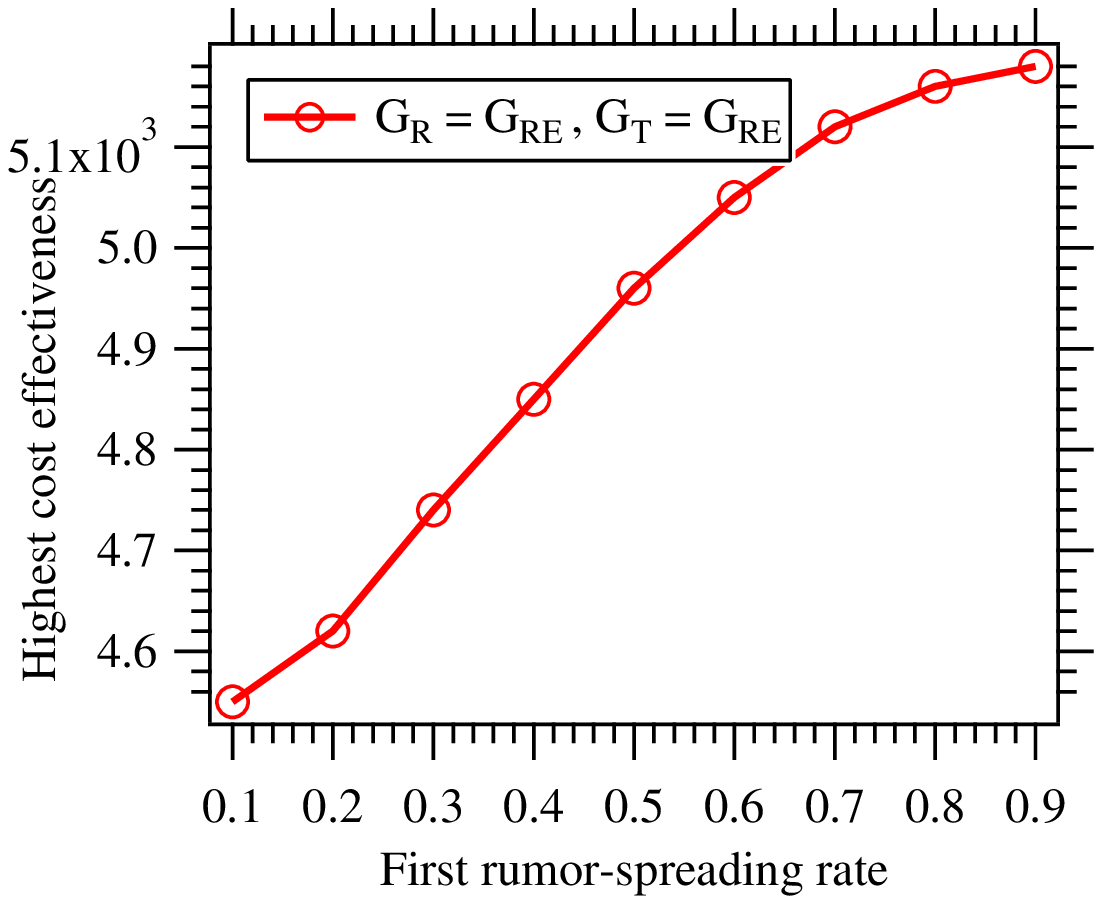}}
  	\caption{The highest cost effectivenesses in Example 3.} 
  	\label{fig:subfig} 
\end{figure}

\begin{expe}
Consider the following RC models:
\begin{tabbing}
  \hspace{1.5cm} $G_R$ \quad\quad\= $G_T$ \quad\quad\= $\beta_1$ \quad\quad\= $\beta_2$ \quad\quad\quad\quad\quad\quad\= $\delta$ \quad\quad\= T \quad\quad\= B \quad\quad\= c$_1$ \quad\quad\= c$_2$ \quad\quad\= $\mathbf{E}^*$\\
  \hspace{1.5cm} $G_{SW}$ \> $G_{SF}$ \> 0.4 \> 0.1/0.2/$\cdots$/0.9 \> 0.1 \> 10 \> 6 \>1 \>2 \> $[0.1, \cdots, 0.1]$\\
  \hspace{1.5cm} $G_{SF}$ \> $G_{SW}$ \> 0.5 \> 0.1/0.2/$\cdots$/0.9 \> 0.2 \> 15 \> 8 \>2 \>3 \> $[0.1, \cdots, 0.1]$\\
  \hspace{1.5cm} $G_{RE}$ \> $G_{RE}$ \> 0.6 \> 0.1/0.2/$\cdots$/0.9 \> 0.3 \> 20 \> 10 \>3 \>4 \> $[0.1, \cdots, 0.1]$
\end{tabbing}
Fig. 7 presents the highest cost effectivenesses for these RC models. It is seen that the highest cost effectiveness is increasing with $\beta_2$. 
\end{expe}
	
\begin{figure}[H]
\centering
	\subfigure[]{\includegraphics[width=0.3\textwidth]{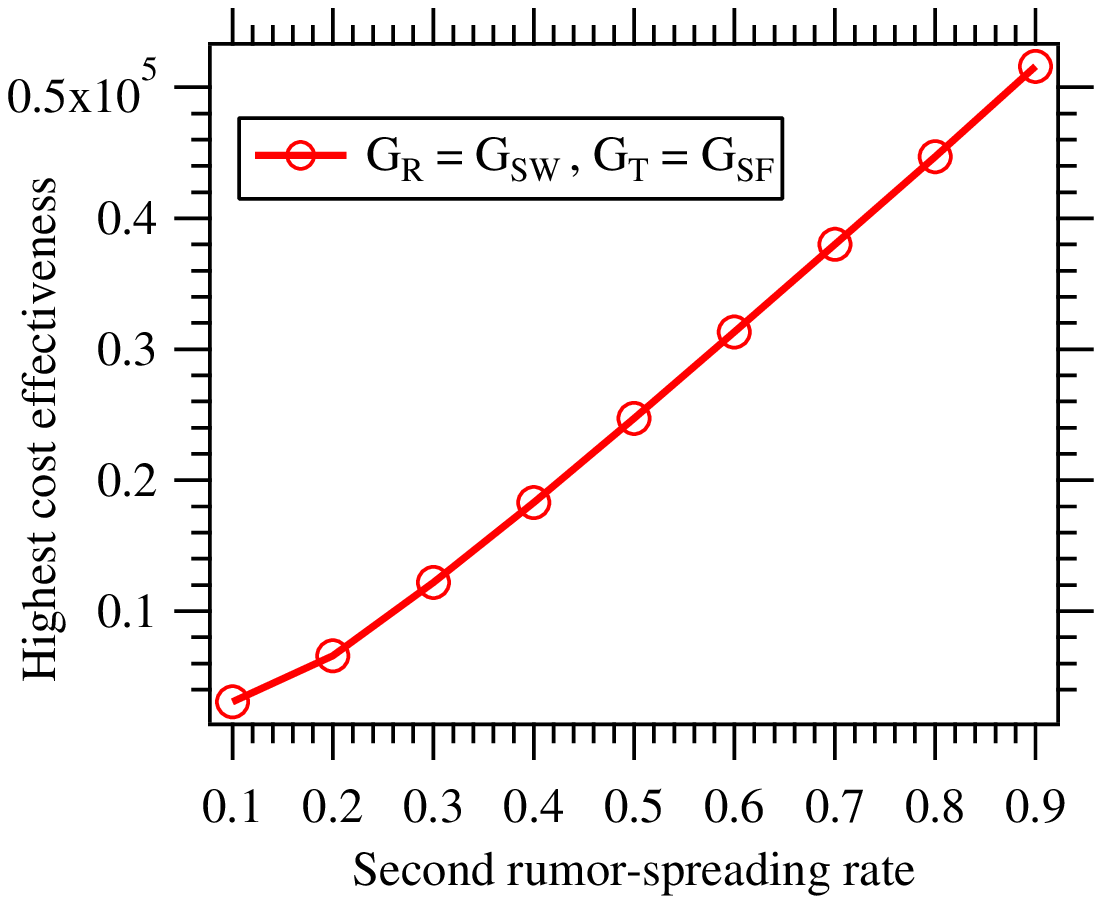}}
	\hspace{.2in}
	\subfigure[]{\includegraphics[width=0.3\textwidth]{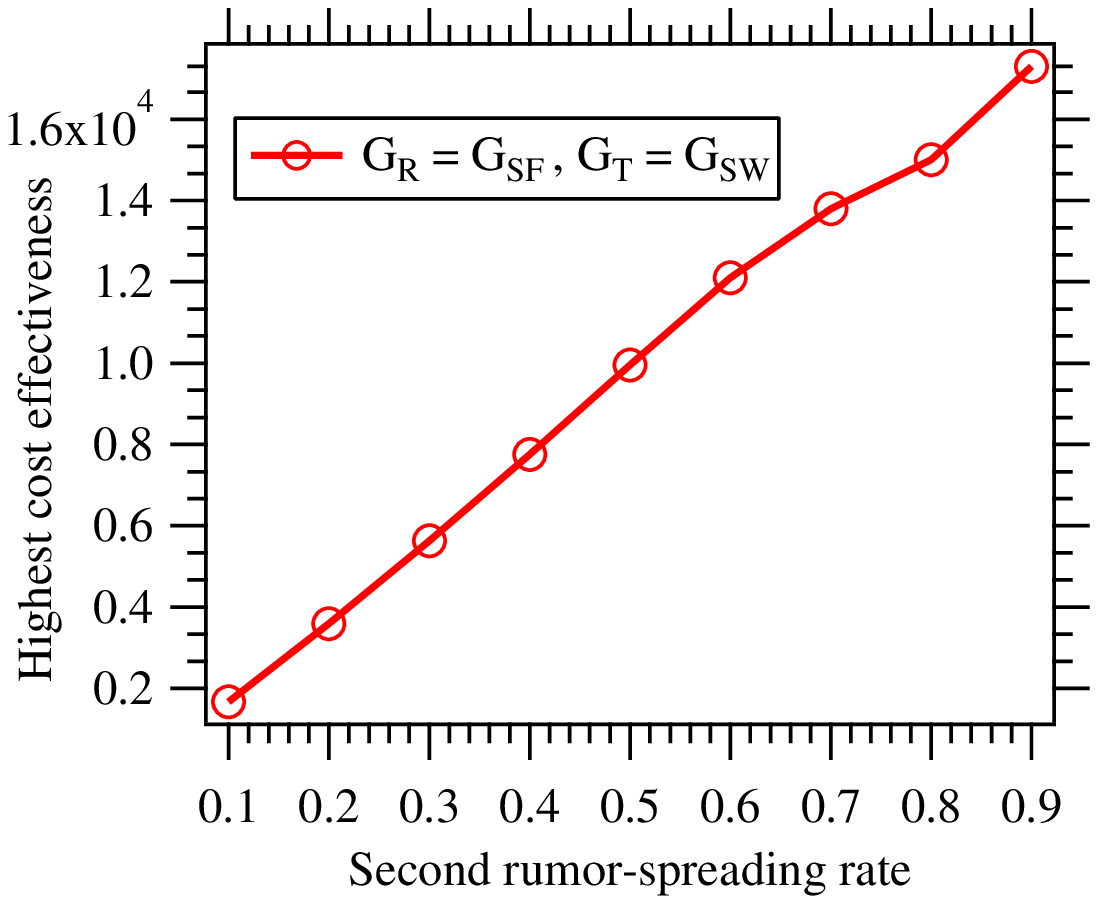}}
	\hspace{.2in}
	\subfigure[]{\includegraphics[width=0.3\textwidth]{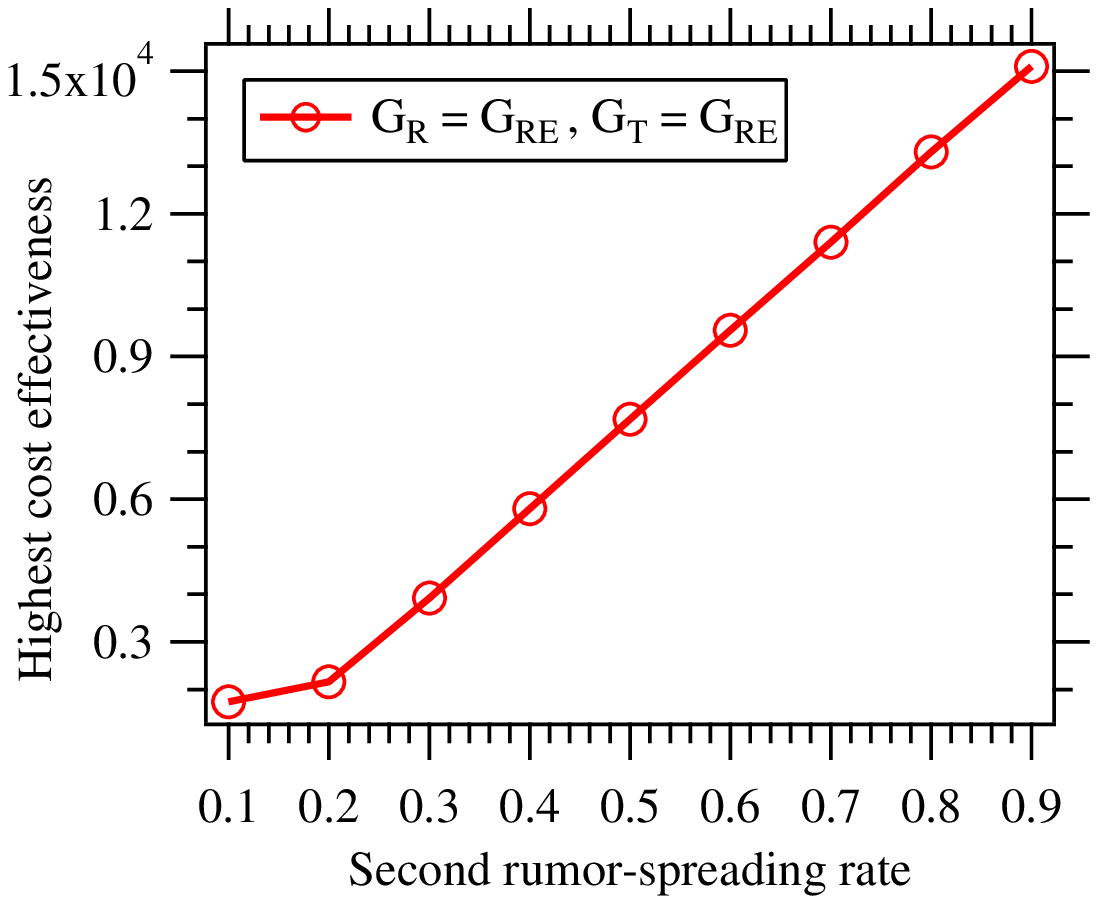}}
	\caption{The highest cost effectivenesses in Example 4.} 
	\label{fig:subfig} 
\end{figure}

Based on these and many similar experiments, we conclude that the highest cost effectiveness always inclines with the two rumor-spreading rates. Therefore, the optimal rumor-containing strategy is especially cost-effective to rumors with higher spreading rates.

\subsection{The influence of the forgetting rate} 

Second, let us consider the influence of the forgetting rate on the highest cost effectiveness.

\begin{expe}
Consider the following RC models:
\begin{tabbing}
  \hspace{1.5cm} $G_R$ \quad\quad\= $G_T$ \quad\quad\= $\beta_1$ \quad\quad\= $\beta_2$ \quad\quad\= $\delta$ \quad\quad\quad\quad\quad\quad\= T \quad\quad\= B \quad\quad\= c$_1$ \quad\quad\= c$_1$ \quad\quad\= $\mathbf{E}^*$\\
  \hspace{1.5cm} $G_{SW}$ \> $G_{SF}$ \> 0.2 \> 0.4 \> 0.1/0.2/$\cdots$/0.9 \> 10 \> 6 \>1 \>2 \> $[0.1, \cdots, 0.1]$\\
  \hspace{1.5cm} $G_{SF}$ \> $G_{SW}$ \> 0.3 \> 0.5 \> 0.1/0.2/$\cdots$/0.9 \> 15 \> 8 \>2 \>3 \> $[0.1, \cdots, 0.1]$\\
  \hspace{1.5cm} $G_{RE}$ \> $G_{RE}$ \> 0.4 \> 0.3 \> 0.1/0.2/$\cdots$/0.9 \> 20 \> 10 \>3 \>4 \> $[0.1, \cdots, 0.1]$
\end{tabbing}
Fig. 8 depicts the highest cost effectivenesses for these RC models. It is seen from this figure that the highest cost effectiveness goes down with $\delta$.
\end{expe}

\begin{figure}[H]
\centering
	\subfigure[]{\includegraphics[width=0.3\textwidth]{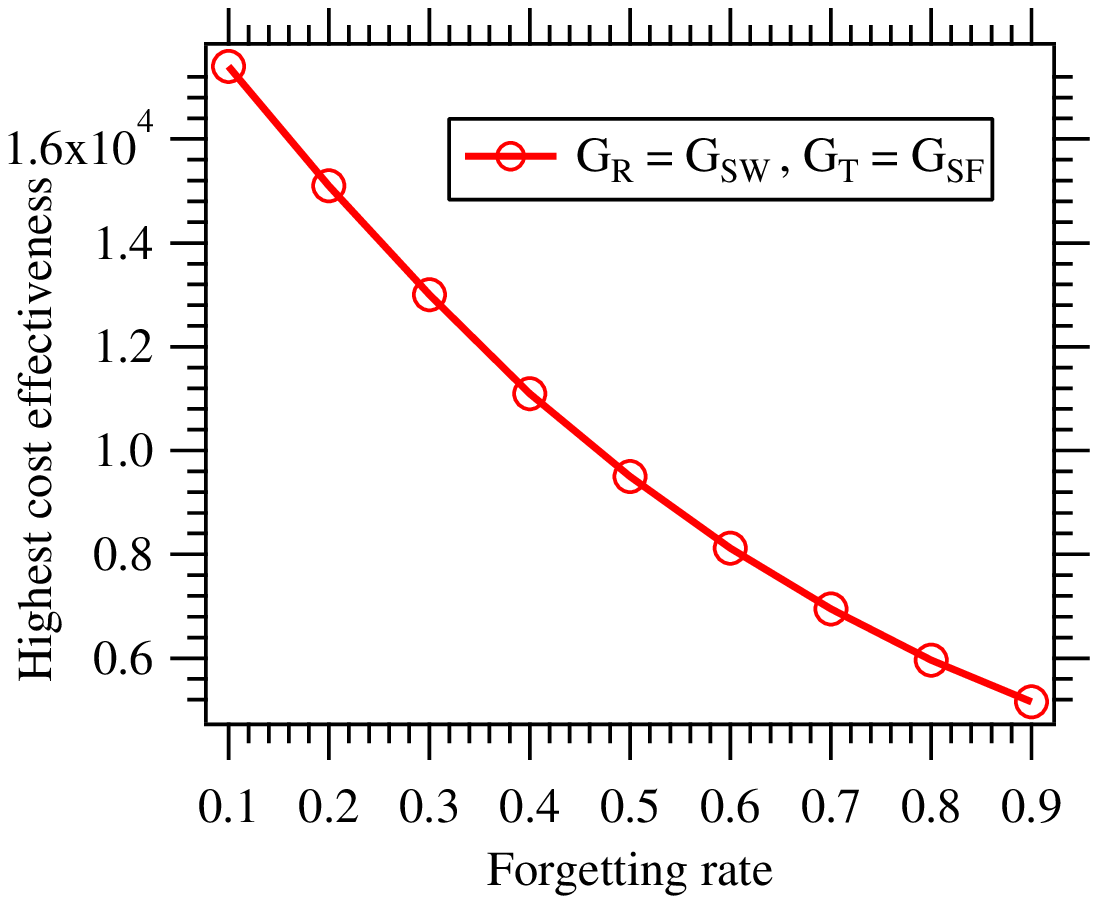}}
	\hspace{.2in}
	\subfigure[]{\includegraphics[width=0.3\textwidth]{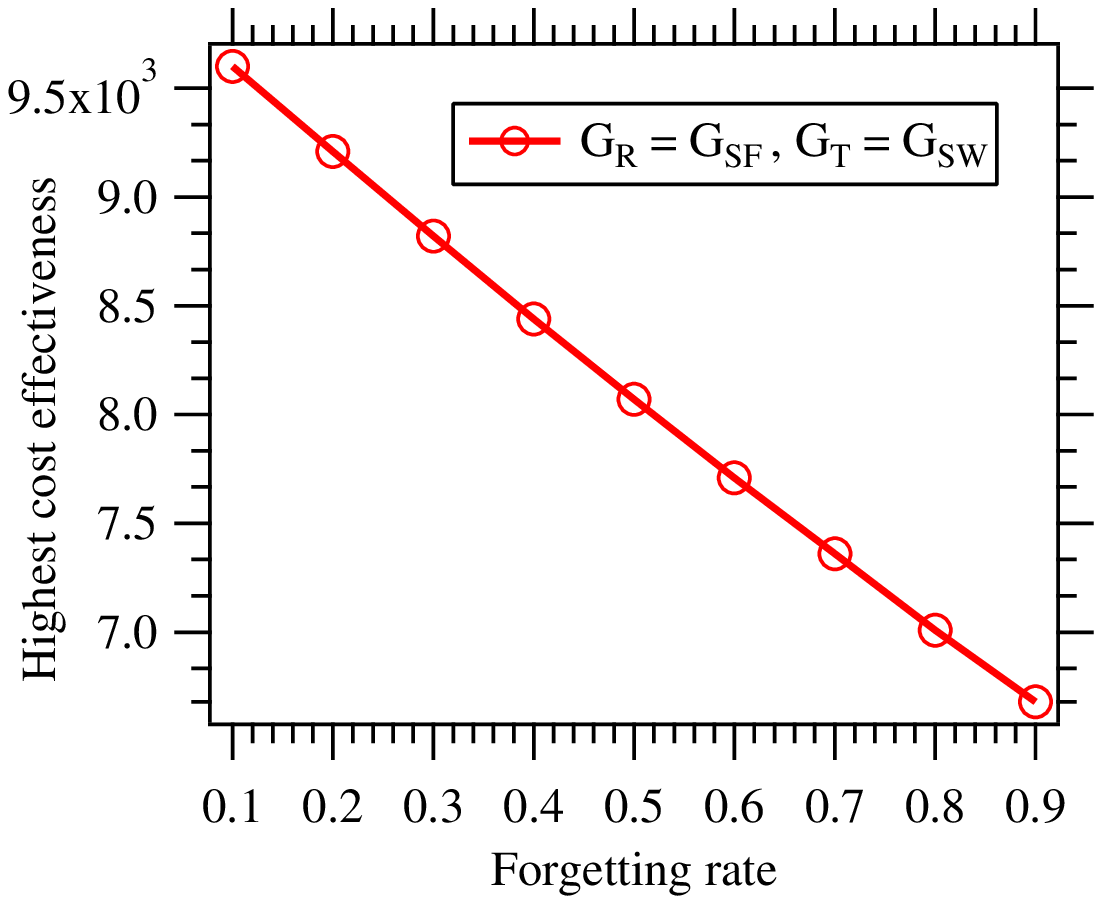}}
	\hspace{.2in}
	\subfigure[]{\includegraphics[width=0.3\textwidth]{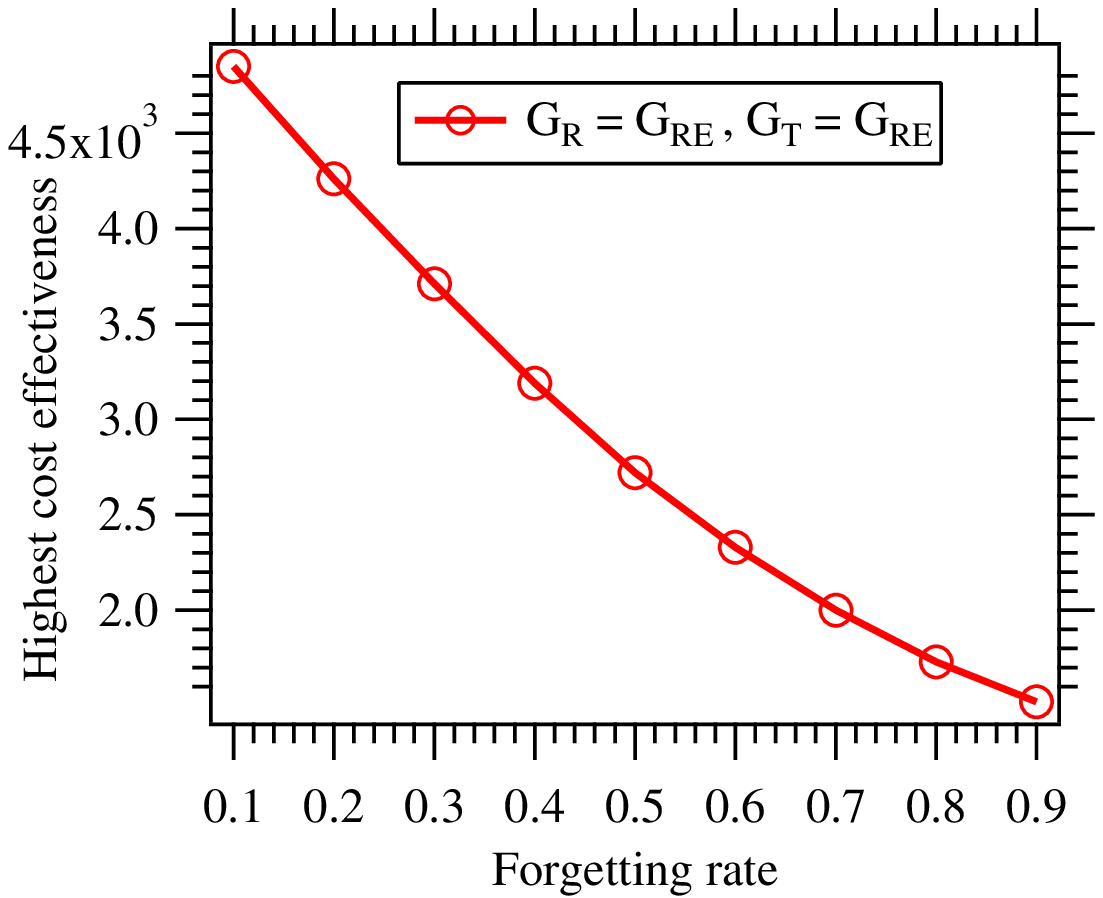}}
  \caption{The highest effectivenesses in Example 5.} 
  \label{fig:subfig} 
\end{figure}

Based on these and many similar experiments, we conclude that the highest cost effectiveness declines with the forgetting rate.

\subsection{The influence of the duration} 

Third, let us inspect the influence of the duration on the highest cost effectiveness.

\begin{expe}
Consider the following RC models:
\begin{tabbing}
  \hspace{1.5cm} $G_R$ \quad\quad\= $G_T$ \quad\quad\= $\beta_1$ \quad\quad\= $\beta_2$ \quad\quad\= $\delta$ \quad\quad\= T \quad\quad\quad\quad\quad\= B \quad\quad\= c$_1$ \quad\quad\= c$_1$ \quad\quad\= $\mathbf{E}^*$\\
  \hspace{1.5cm} $G_{SW}$ \> $G_{SF}$ \> 0.3 \>0.5 \> 0.1 \>10/12/$\cdots$/30 \> 6 \>1 \>2 \> $[0.1, \cdots, 0.1]$\\
  \hspace{1.5cm} $G_{SF}$ \> $G_{SW}$ \> 0.4 \> 0.6 \>0.2 \>10/12/$\cdots$/30 \> 8 \>2 \>3 \> $[0.1, \cdots, 0.1]$\\
  \hspace{1.5cm} $G_{RE}$ \> $G_{RE}$ \> 0.6 \> 0.4 \> 0.3 \>10/12/$\cdots$/30 \> 10 \>3 \>4 \> $[0.1, \cdots, 0.1]$
	\end{tabbing}
Fig. 9 exhibits the highest cost effectivenesses for these RC models. It is seen that the highest cost effectiveness goes up with $T$.
\end{expe}

\begin{figure}[H]
	\centering
	\subfigure[]{\includegraphics[width=0.3\textwidth]{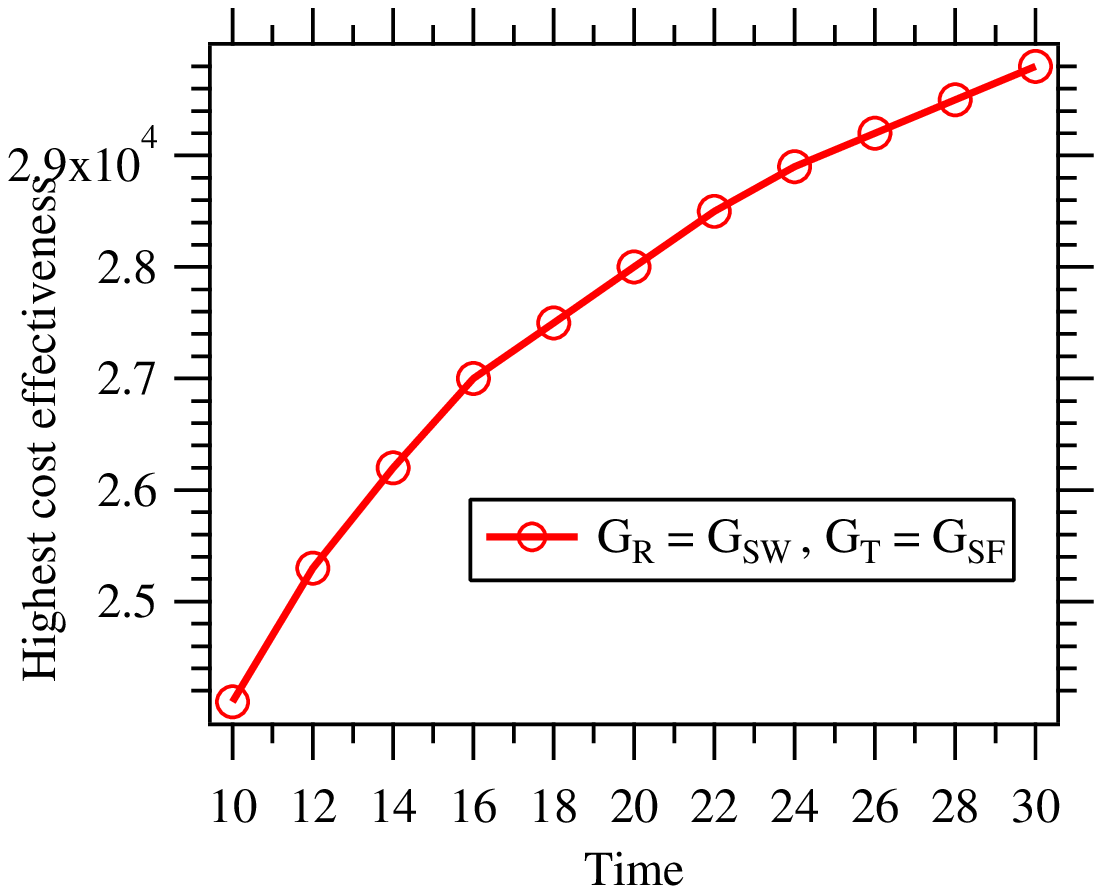}}
	\hspace{.2in}
	\subfigure[]{\includegraphics[width=0.3\textwidth]{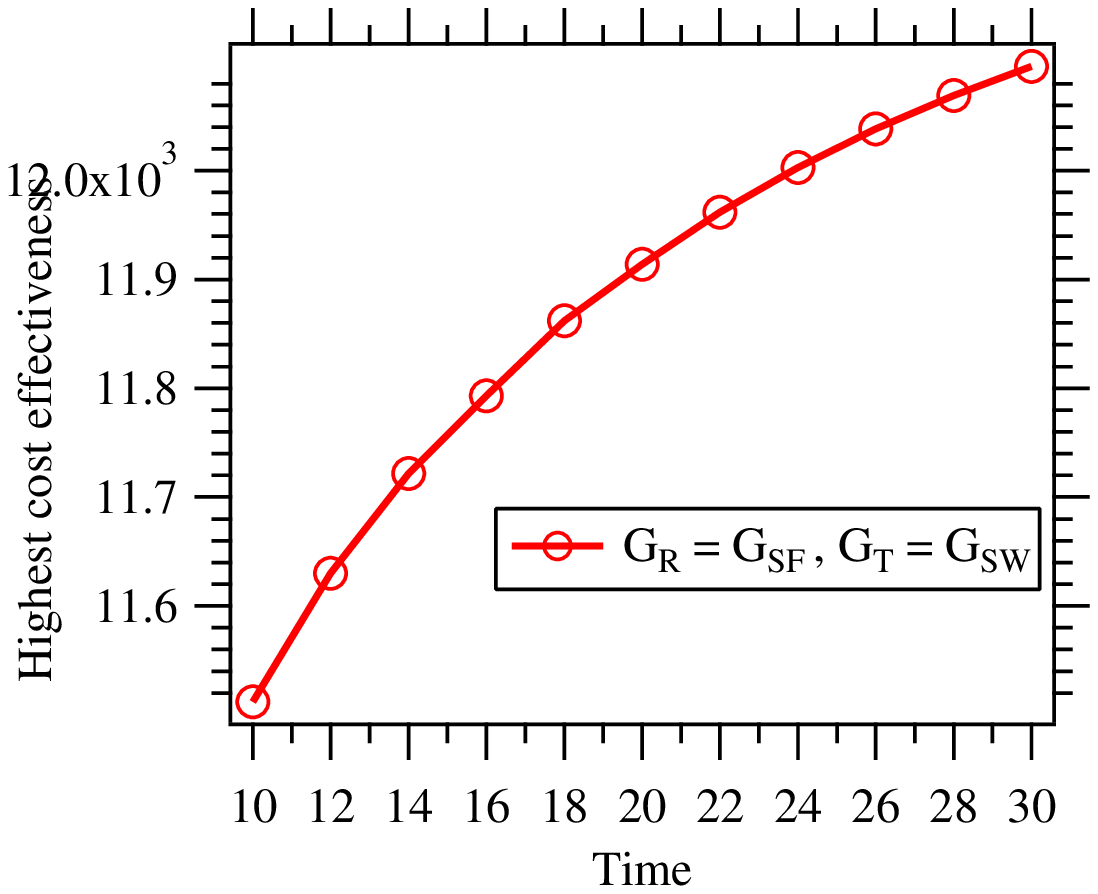}}
	\hspace{.2in}
	\subfigure[]{\includegraphics[width=0.3\textwidth]{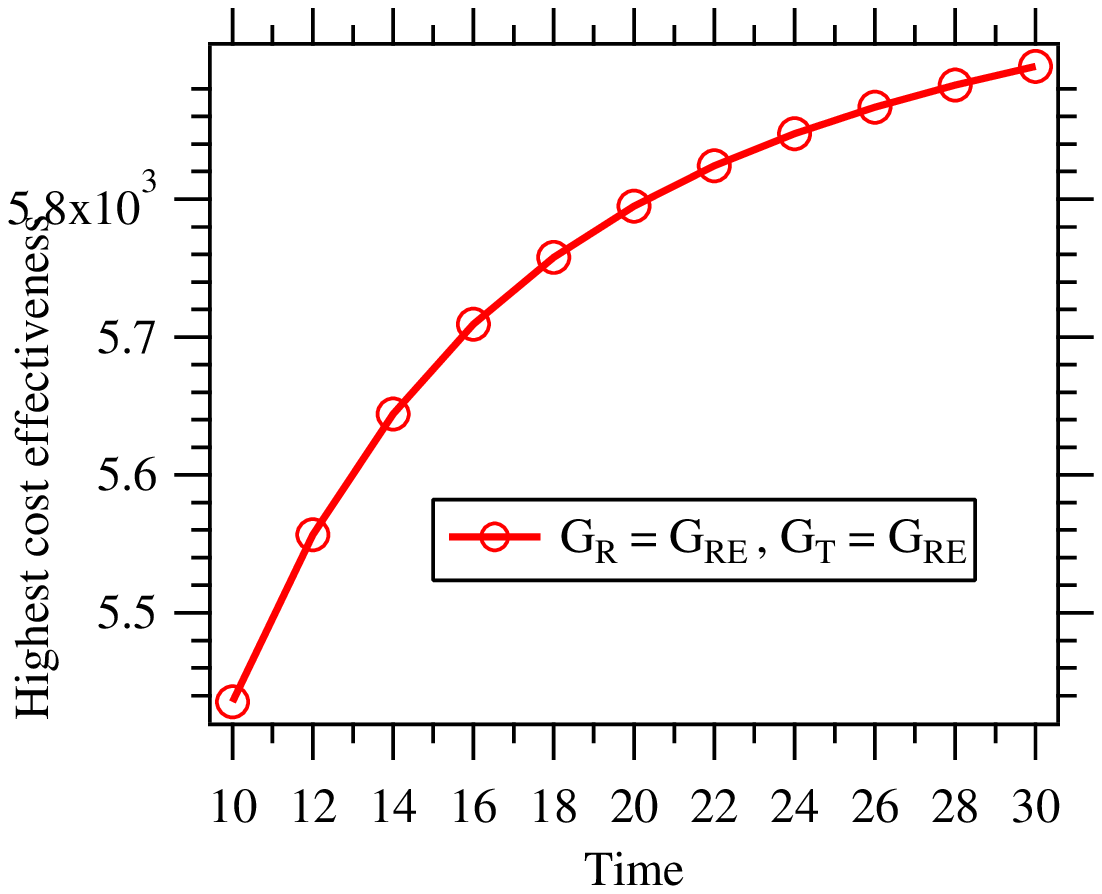}}	
	\caption{The highest cost effectivenesses in Example 6.} 
	\label{fig:subfig} 
\end{figure}

From these and many similar experiments, we conclude that the highest cost effectiveness always inclines with the duration. Therefore, with the extension of the duration, the optimal rumor-containing strategy becomes increasingly cost-effective.

\subsection{The influence of the rumor-containing budget per unit time} 

Next, we investigate the influence of the rumor-containing budget per unit time on the highest cost effectiveness.

\begin{expe}
Consider the following RC models:
\begin{tabbing}
  \hspace{1.5cm} $G_R$ \quad\quad\= $G_T$ \quad\quad\= $\beta_1$ \quad\quad\= $\beta_2$ \quad\quad\= $\delta$ \quad\quad\= T \quad\quad\= B \quad\quad\quad\quad\quad\= c$_1$ \quad\quad\= c$_2$ \quad\quad\= $\mathbf{E}^*$\\
  \hspace{1.5cm} $G_{SW}$ \> $G_{SF}$ \> 0.4 \> 0.2\> 0.8 \>60 \> 2/4/$\cdots$/18 \> 7 \>9 \> $[0.1, \cdots, 0.1]$\\
  \hspace{1.5cm} $G_{SF}$ \> $G_{SW}$ \> 0.8 \> 0.6\> 0.8 \>70 \> 2/4/$\cdots$/18 \> 8 \>8 \> $[0.1, \cdots, 0.1]$\\
  \hspace{1.5cm} $G_{RE}$ \> $G_{RE}$ \> 0.9 \> 0.2\> 0.5 \>65 \> 2/4/$\cdots$/18 \> 5 \>7 \> $[0.1, \cdots, 0.1]$
\end{tabbing}
Fig. 10 plots the highest cost effectivenesses for these RC models. It is seen that, with the increase of $B$, the highest cost effectiveness first goes up then goes down.
\end{expe}

\begin{figure}[H]
\centering
	\subfigure[]{\includegraphics[width=0.3\textwidth]{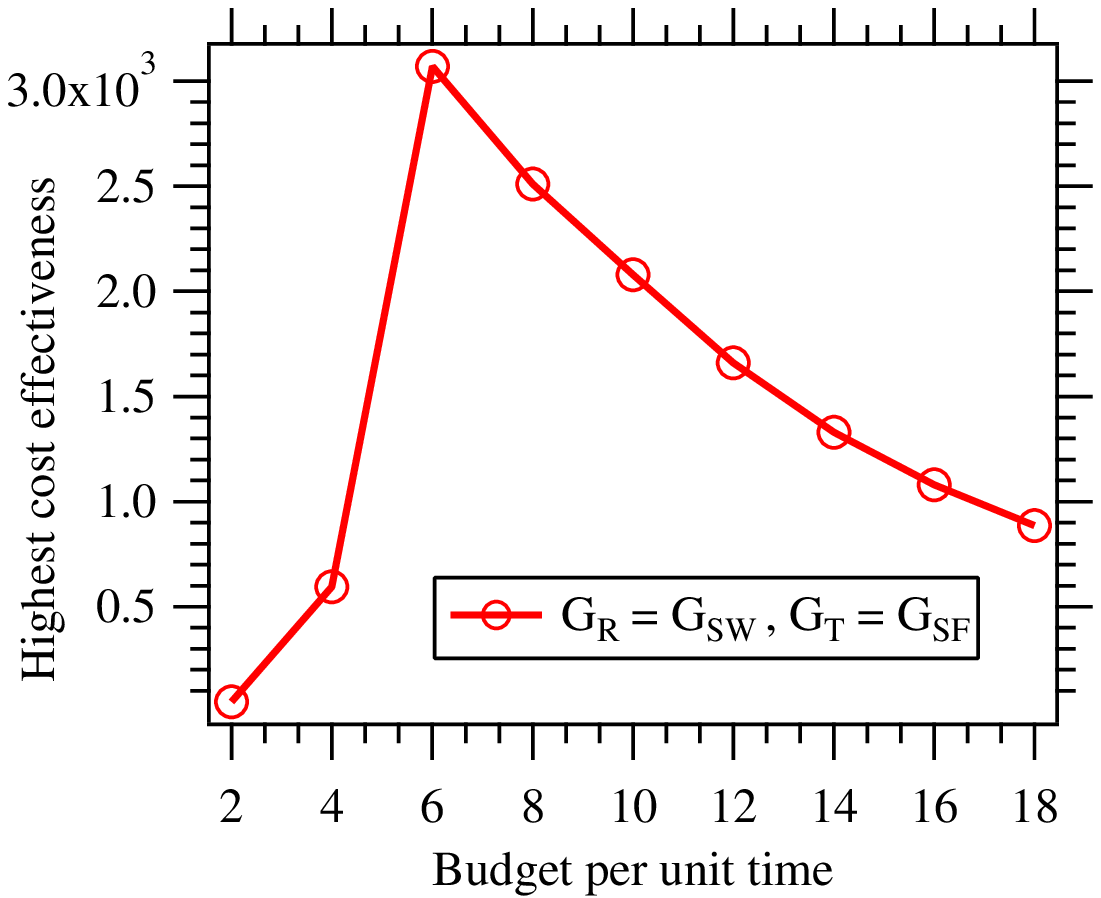}}
	\hspace{.2in}
	\subfigure[]{\includegraphics[width=0.3\textwidth]{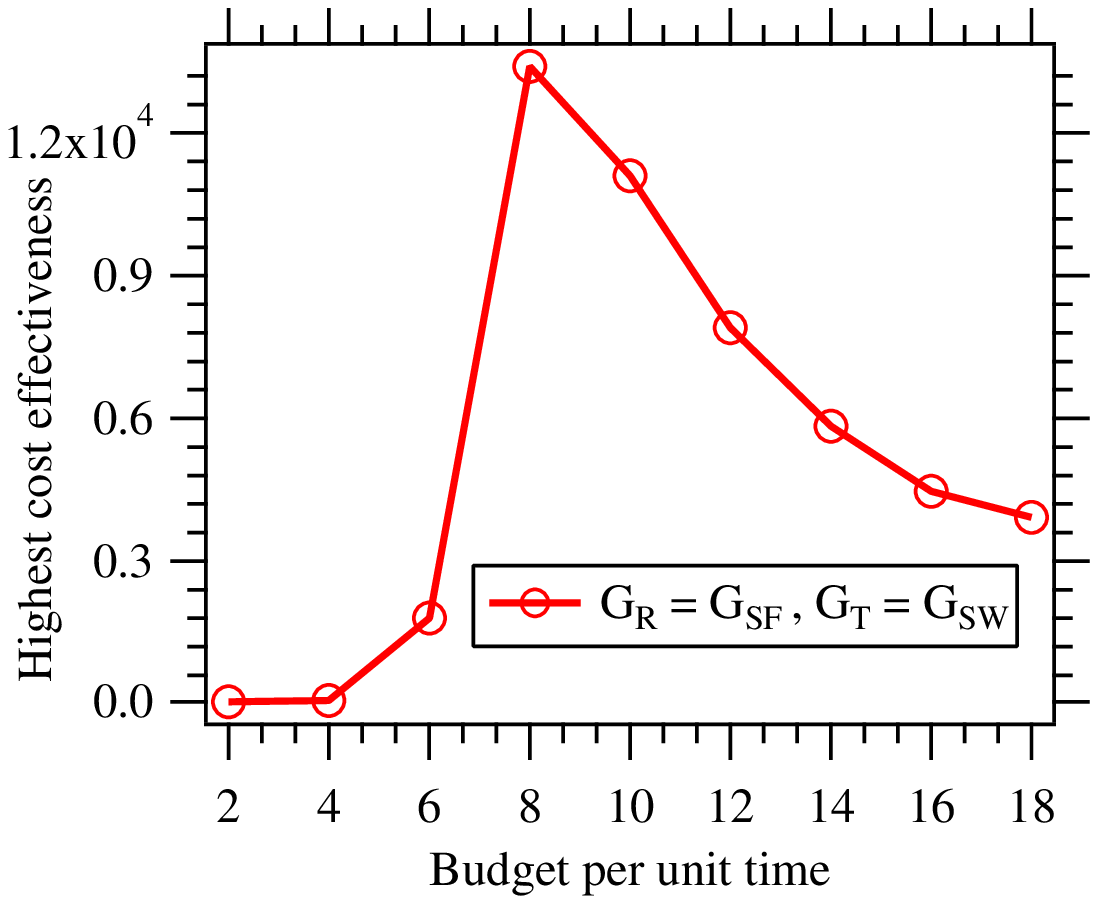}}
	\hspace{.2in}
	\subfigure[]{\includegraphics[width=0.3\textwidth]{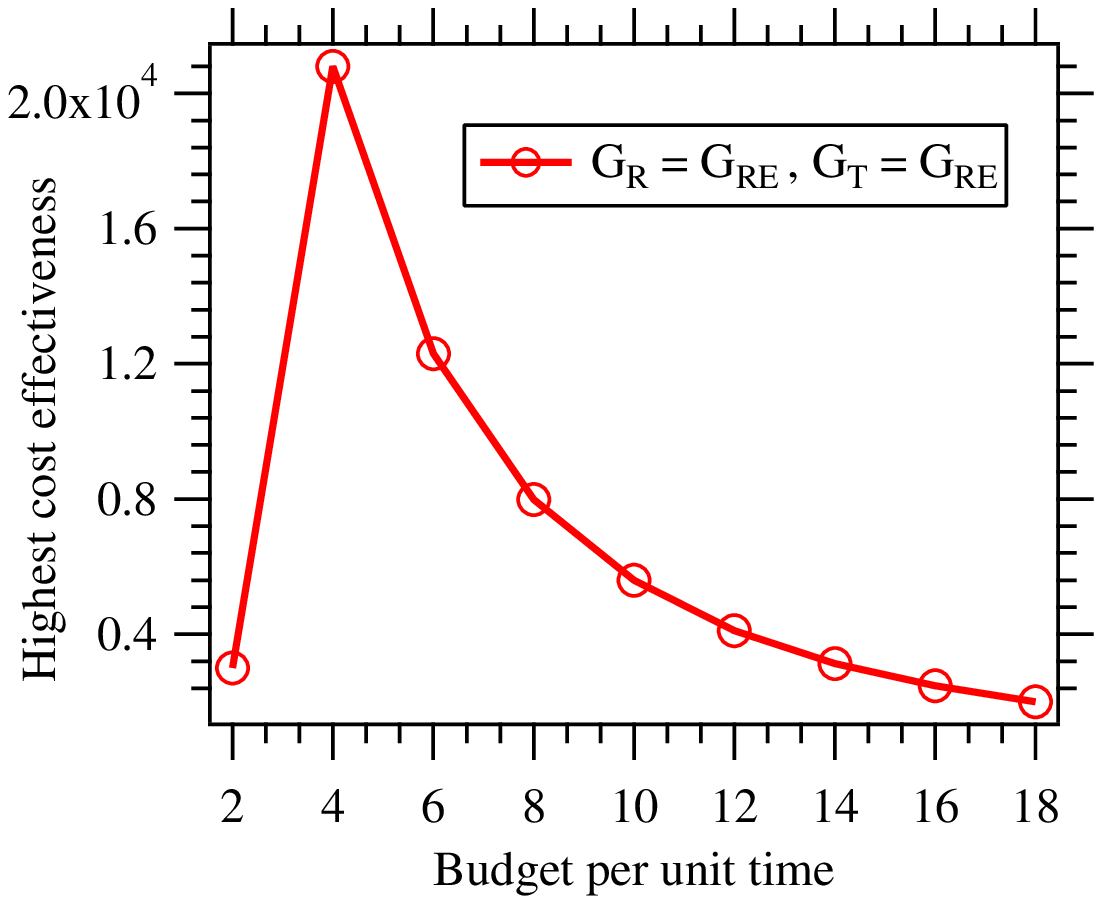}}
  \caption{The highest cost effectivenesses in Example 7.} 
  \label{fig:subfig} 
\end{figure}

From these and many similar experiments, we conclude that, with the rumor-containing budget per unit time, the highest cost effectiveness always first inclines then declines. This phenomenon is interesting, because it implies that a medium-scale rumor-containing budget per unit time would achieve the highest cost effectiveness.

\subsection{The influence of the two truth-spreading cost coefficients} 

Finally, we study the influence of the two truth-spreading costs on the highest cost effectiveness.

\begin{expe}
Consider the following RC models:
\begin{tabbing}
  \hspace{1.5cm} $G_R$ \quad\quad\= $G_T$ \quad\quad\= $\beta_1$ \quad\quad\= $\beta_2$ \quad\quad\= $\delta$ \quad\quad\= T \quad\quad\= B \quad\quad\= c$_1$ \quad\quad\quad\quad\= c$_2$ \quad\quad\= $\mathbf{E}^*$\\
  \hspace{1.5cm} $G_{SW}$ \> $G_{SF}$ \> 0.5 \> 0.4 \> 0.8 \> 50 \> 2 \> 1/2/$\cdots$/9 \>5 \> $[0.1, \cdots, 0.1]$\\
  \hspace{1.5cm} $G_{SF}$ \> $G_{SW}$ \> 0.3 \> 0.9 \> 0.6 \> 30 \> 12 \>1/2/$\cdots$/9 \>5 \> $[0.1, \cdots, 0.1]$\\
  \hspace{1.5cm} $G_{RE}$ \> $G_{RE}$ \> 0.4 \> 0.8 \> 0.2 \> 50 \> 14 \>1/2/$\cdots$/9 \>8 \> $[0.1, \cdots, 0.1]$
\end{tabbing}
Fig. 11 plots the highest cost effectivenesses for these RC models. It is seen that the highest cost effectiveness is decreasing with $c_1$. 
\end{expe}

\begin{figure}[H]
	\centering
	\subfigure[]{\includegraphics[width=0.3\textwidth]{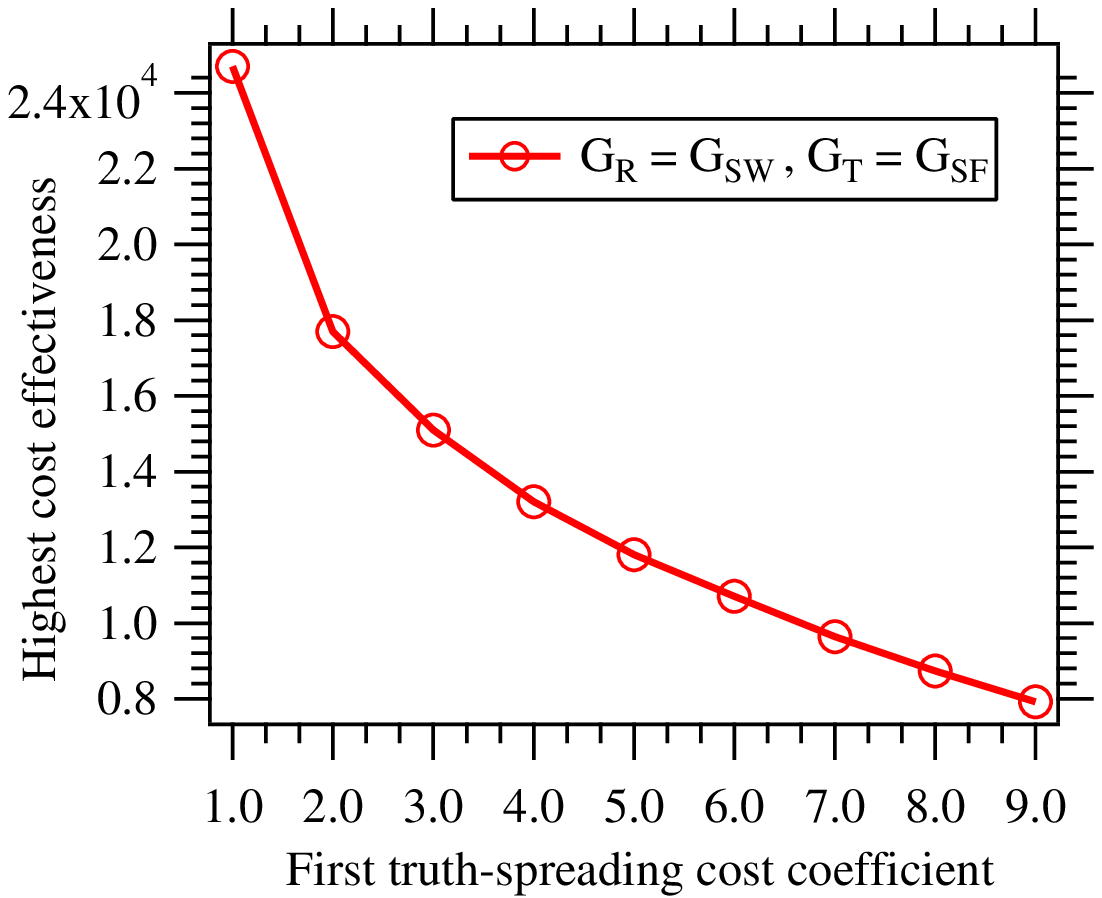}}
	\hspace{.2in}
	\subfigure[]{\includegraphics[width=0.3\textwidth]{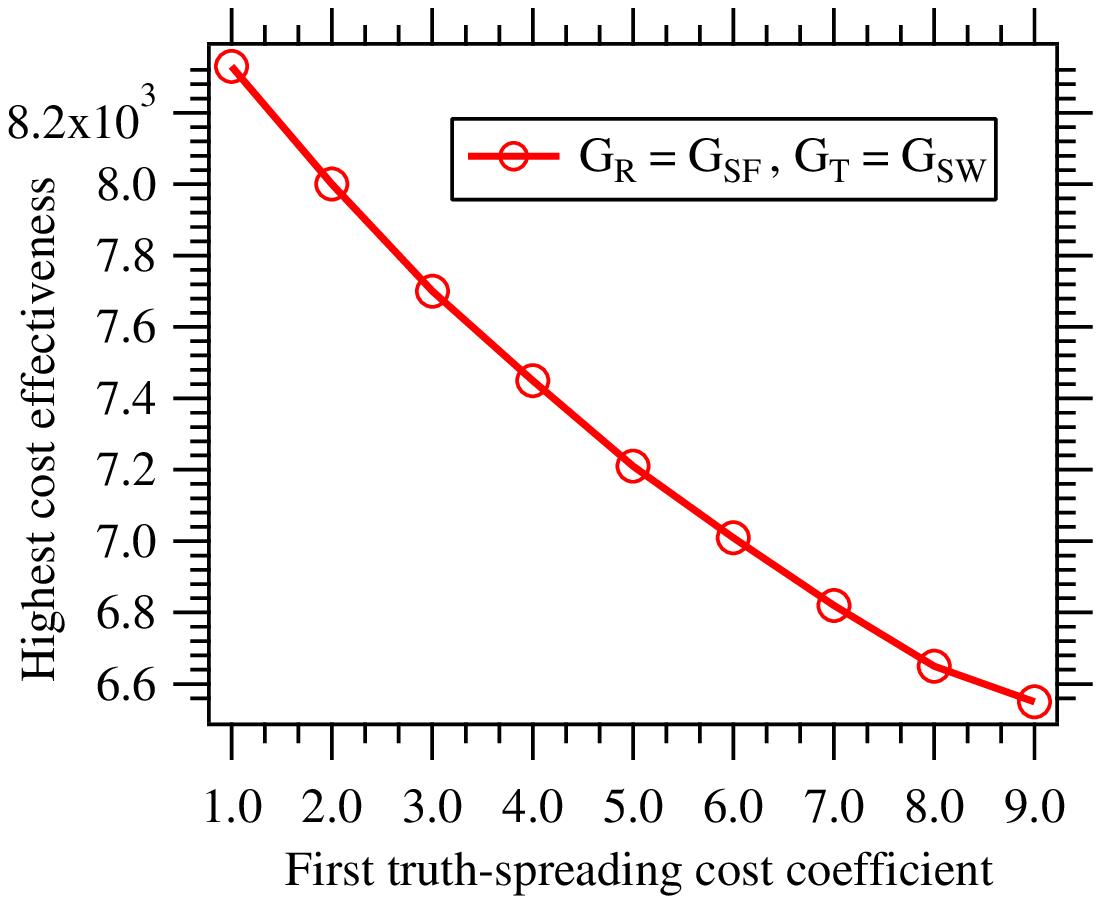}}
	\hspace{.2in}
	\subfigure[]{\includegraphics[width=0.3\textwidth]{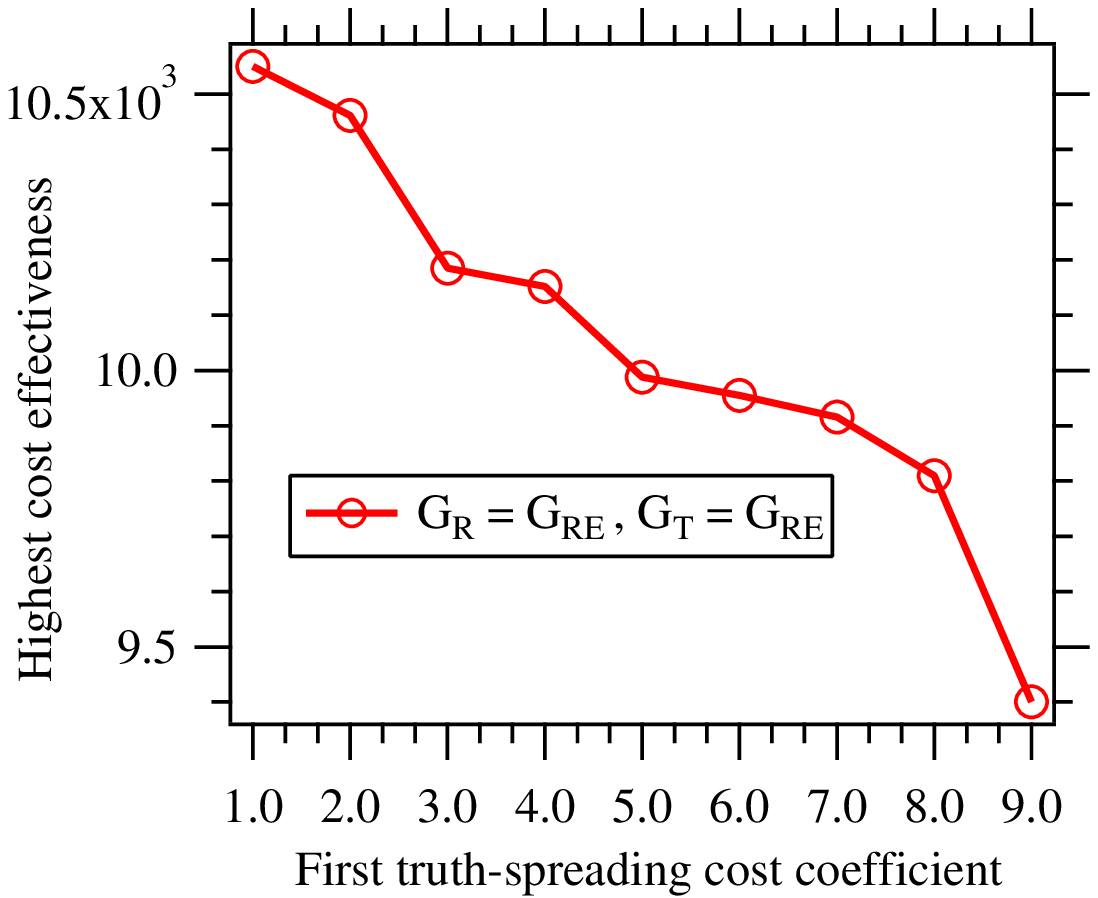}}
	\caption{The highest cost effectivenesses in Example 8.} 
	\label{fig:subfig} 
\end{figure}

\begin{expe}
Consider the following RC models:
\begin{tabbing}
  \hspace{1.5cm} $G_R$ \quad\quad\= $G_T$ \quad\quad\= $\beta_1$ \quad\quad\= $\beta_2$ \quad\quad\= $\delta$ \quad\quad\= T \quad\quad\= B \quad\quad\= c$_1$ \quad\quad\= c$_2$ \quad\quad\quad\quad\quad\= $\mathbf{E}^*$\\
  \hspace{1.5cm} $G_{SW}$ \> $G_{SF}$ \> 0.7 \> 0.3 \> 0.9 \> 55 \> 12\>2 \>1/2/$\cdots$/9 \> $[0.1, \cdots, 0.1]$\\
  \hspace{1.5cm} $G_{SF}$ \> $G_{SW}$ \> 0.3 \> 0.1 \> 0.2 \> 30 \> 16\>7 \>1/2/$\cdots$/9 \> $[0.1, \cdots, 0.1]$\\
  \hspace{1.5cm} $G_{RE}$ \> $G_{RE}$ \> 0.4 \> 0.2 \> 0.4 \> 50 \> 4 \>5 \>1/2/$\cdots$/9 \> $[0.1, \cdots, 0.1]$
\end{tabbing}
Fig. 12 plots the highest cost effectivenesses for these RC models. It is seen that the highest cost effectiveness is increasing with $c_2$. 
\end{expe}

\begin{figure}[H]
	\centering
	\subfigure[]{\includegraphics[width=0.3\textwidth]{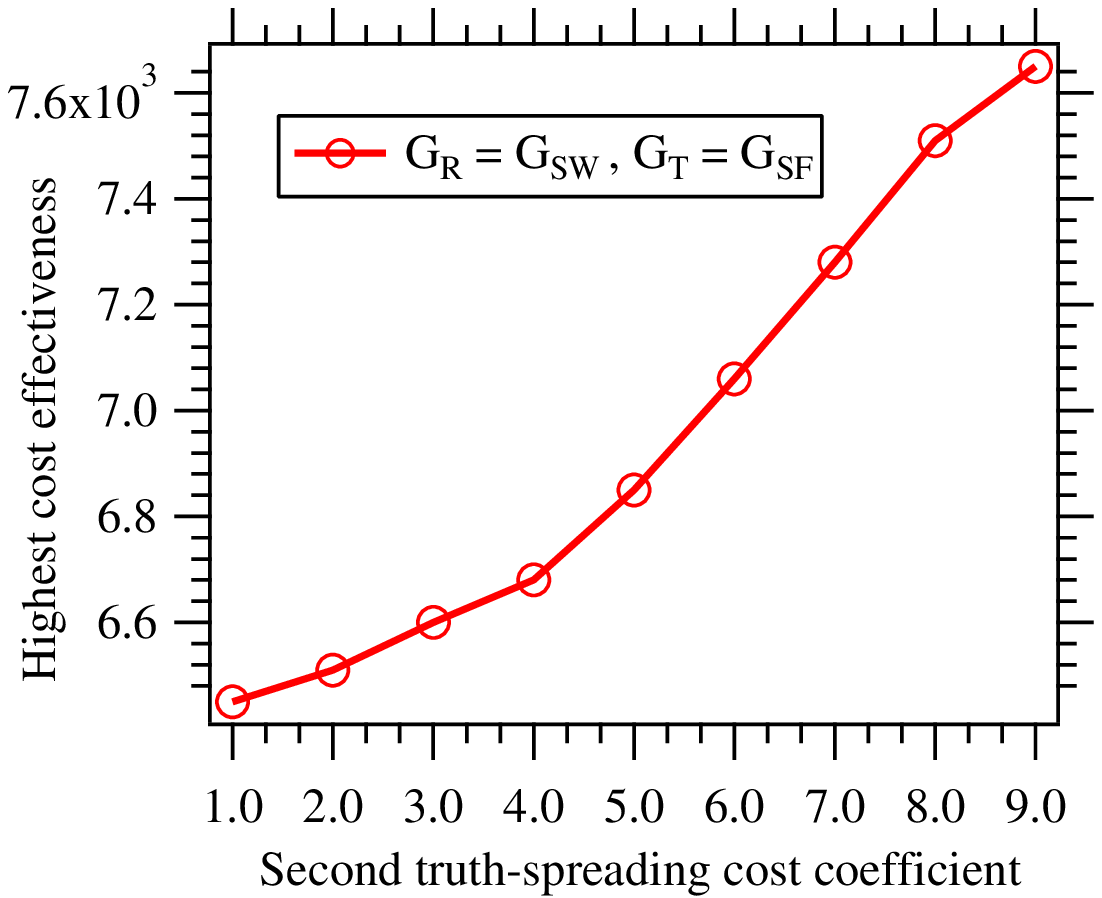}}
	\hspace{.2in}
	\subfigure[]{\includegraphics[width=0.3\textwidth]{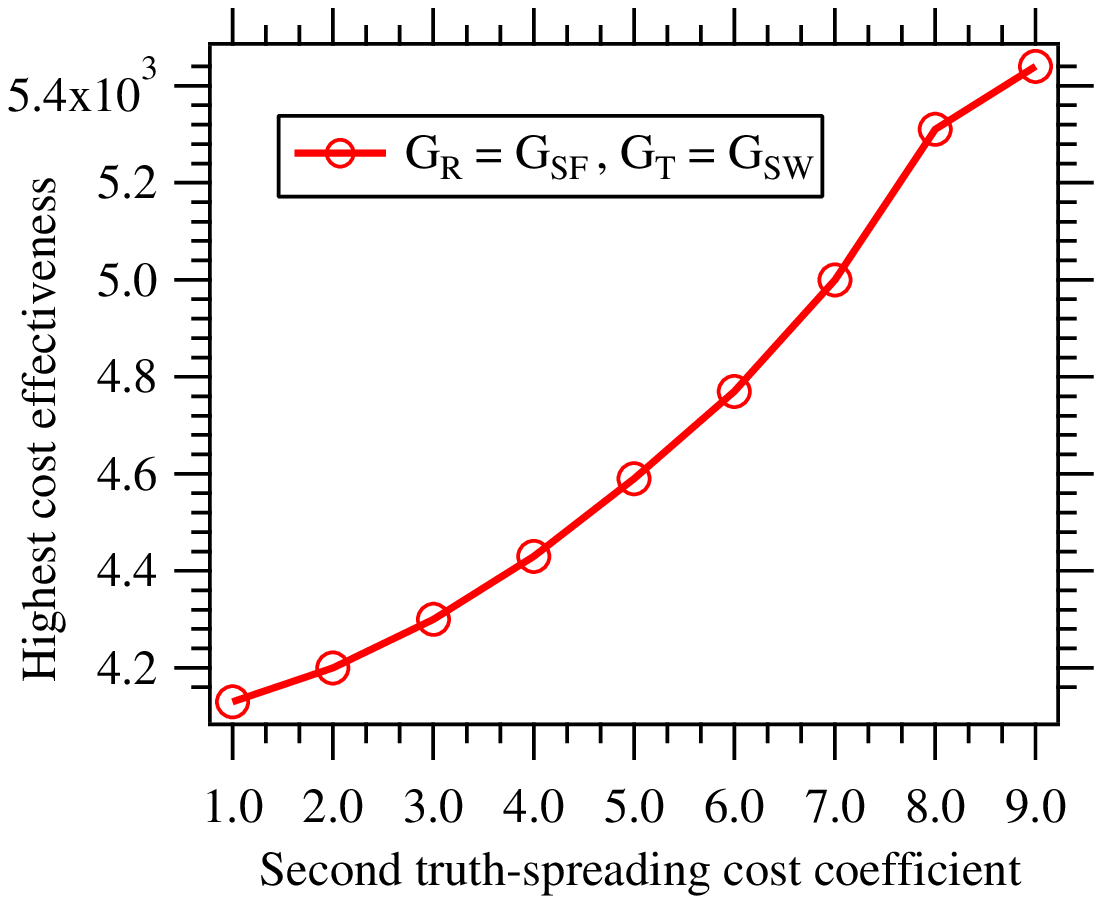}}
	\hspace{.2in}
	\subfigure[]{\includegraphics[width=0.3\textwidth]{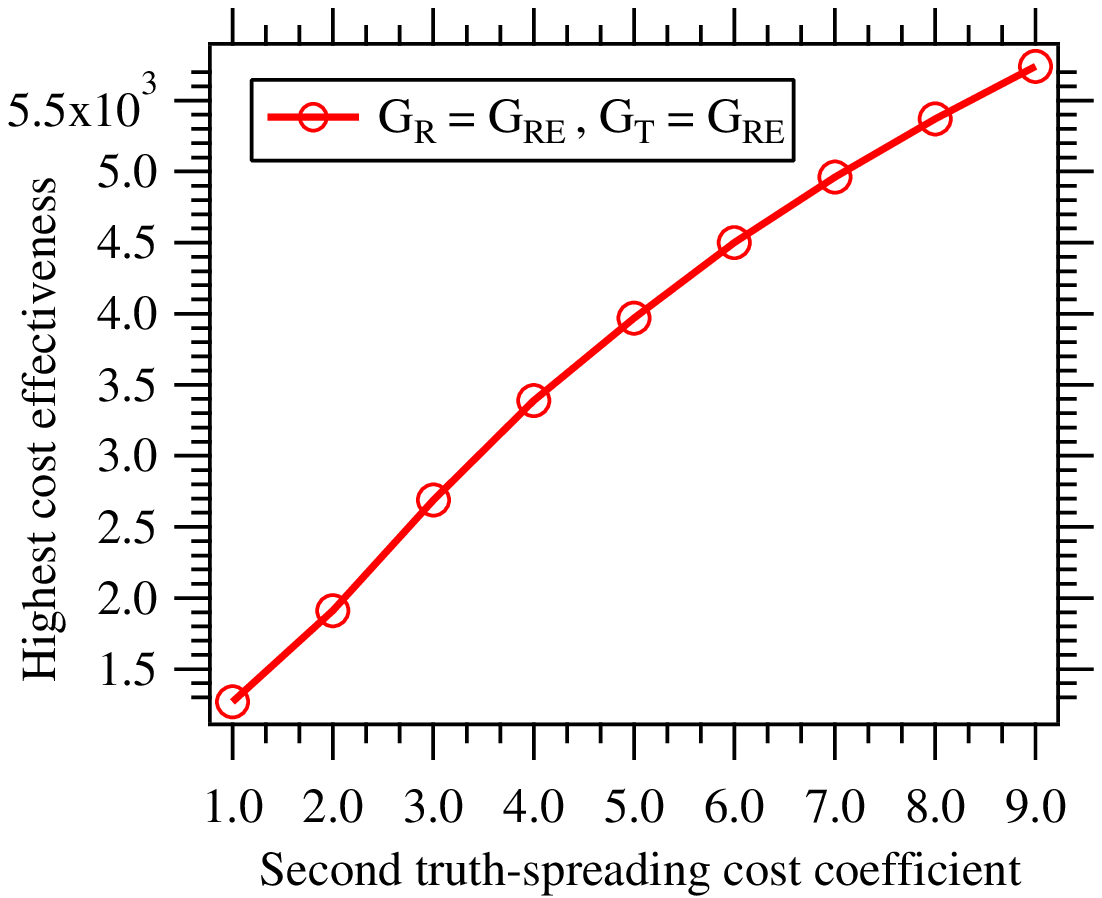}}
	\caption{The highest cost effectivenesses in Example 9.} 
	\label{fig:subfig} 
\end{figure}

Based on these and many similar experiments, we conclude that the highest cost effectiveness always declines with the first truth-spreading cost coefficient and inclines with the second truth-spreading cost coefficient.

\section{Concluding remarks}
	
This paper has addressed the problem of how to inhibit a false rumor. Based on a novel rumor-truth mixed spreading model, the design problem has been modeled as an optimization problem (the RC model), with the goal of finding the most effective rumor-containing strategy subject to a limited rumor-containing budget. Some optimal rumor-containing strategies have been obtained by solving the RC model. Additionally, we have revealed the influence of different factors, including the two rumor-spreading rates, the forgetting rate, the duration, the rumor-containing budget per unit time, and the two truth-spreading cost coefficients, on the highest cost effectiveness of a RC model.
	
Toward this direction there are some interesting research topics. This work builds on the premise that all nodes are homogeneous. It is valuable to study the rumor-containing problem based on the more realistic node heterogeneity assumption \cite{GuoQ2016, ChenM2016, WangCJ2017}. The rumor suppression under rumor spreading models with diffusion terms is an interesting topic \cite{ZhuLH2016, ZhuLH2017}. Usually, rumors may spread along various channels. Hence, it is of practical importance to study the rumor-containing problem on interconnected networks \cite{Dickison2012, WangHJ2013}. In this work, all the rumor-containing strategies are assumed to be fixed. In real-world applications, a rumor-containing strategy may well be adjusted flexibly to achieve a higher cost effectiveness. In this context, a more cost-effective rumor-containing strategy should be figured out in the framework of optimal control theory \cite{Kirk2004, YangLX2016, Nowzari2016, ZhangTR2017, BiJC2017}. Even further, both the rumor-spreading strategy and the rumor-containing strategy may vary over time. Game theory provides a proper framework for studying the rumor-containing problem in this background \cite{Osborne2003, Alpcan2010, Manshaei2013}.

\section*{Acknowledgments}
	
	
The authors are grateful to the five anonymous reviewers and the editor for their valuable comments and suggestions. This work is supported by National Natural Science Foundation of China (Grant No. 61572006), Sci-Tech Support Program of China (Grant No. 2015BAF05B03) and Fundamental Research Funds for the Central Universities (Grant No. 106112014CDJZR008823).

\section*{References}
	\bibliographystyle{elsarticle-num}
	\biboptions{numbers,sort&compress}
	\bibliography{<your-bib-database>}
	
\end{document}